\documentclass[dvipsnames,aps,prx,nobibnotes,twocolumn,superscriptaddress,longbibliography]{revtex4-2}

\usepackage[T1]{fontenc}
\usepackage[english]{babel}
\usepackage[utf8]{inputenc}

\usepackage[unicode=true,pdfusetitle,bookmarks=true,bookmarksnumbered=false,bookmarksopen=false,
 breaklinks=false,pdfborder={0 0 0},pdfborderstyle={},backref=false,colorlinks=true]
 {hyperref}
 \hypersetup{
 citecolor=blue,
 urlcolor=blue,
 linkcolor=blue}

\usepackage{array}
\usepackage{wrapfig}
\usepackage{float}
\usepackage{units}
\usepackage{amsmath}
\usepackage{cancel}
\usepackage{mathrsfs}
\usepackage{amssymb}
\usepackage{graphicx}
\usepackage{stackrel}
\usepackage{babel}
\usepackage{xparse}
\usepackage{soul}
\usepackage[final]{pdfpages}
\usepackage{comment}
\usepackage{tcolorbox}
\usepackage[final]{pdfpages}

\newcommand{\new}[1]{{\color{black} {#1}}}

\makeatletter
\let\oldsection\section 
\RenewDocumentCommand{\section}{s o m}{%
  \IfBooleanTF{#1}
    {\oldsection*{\MakeUppercase{#3}}} 
    {\IfValueTF{#2}
       {\oldsection[#2]{\MakeUppercase{#3}}} 
       {\oldsection{\MakeUppercase{#3}}}     
    }%
}
\AtBeginDocument{\let\LS@rot\@undefined}
\makeatother

\begin{document}
\title{Geometric Criticality in Scale-Invariant Networks}
\author{Lorenzo Lucarini}
\affiliation{Physics Department and INFN, University of Rome Tor Vergata, 00133 Rome, Italy}
\affiliation{Enrico Fermi Research Center (CREF), 00184, Rome, Italy}
\author{Giulio Cimini}
\affiliation{Physics Department and INFN, University of Rome Tor Vergata, 00133 Rome, Italy}
\affiliation{Enrico Fermi Research Center (CREF), 00184, Rome, Italy}
\author{Pablo Villegas}
\email{pablo.villegas@cref.it}
\affiliation{Enrico Fermi Research Center (CREF), 00184, Rome, Italy}
\affiliation{Instituto Carlos I de F\'isica Te\'orica y Computacional, Universidad de Granada, 18071 Granada, Spain}
\begin{abstract}
Dimension in physical systems determines universal properties at criticality. Yet, the impact of structural perturbations on dimensionality remains largely unexplored. Here, we characterize the attraction basins of structural fixed points in scale-invariant networks from a renormalization group perspective, demonstrating that basin stability connects to a structural phase transition. This topology-dependent effect, which we term {\em geometric criticality}, triggers a geometric breakdown \new{hitherto} unknown, which induces non-trivial fractal dimensions and unveils hidden \new{LRG} flows toward unstable structural fixed points. Our systematic study of how networks and lattices respond to disorder paves the way for future analysis of non-ergodic behavior induced by quenched disorder.
\end{abstract}

\maketitle
Topological shortcuts and structural sparsity shape complex dynamics across a broad spectrum of physical systems \cite{Betzel2018, Martin2015, Simonov2020}. Shortcuts are key, for instance, in shaping brain functionality \cite{SpornsBook} or in improving the speed, efficiency, robustness, and functional diversity across neural networks \cite{Kaiser2006,Bullmore2009,Betzel2017}. Similar concepts pervade other areas, such as wormholes in general relativity \cite{Wormholes} or wormhole attacks in wireless sensor networks \cite{Chen2014}. In contrast, sparsity -- for example, selective bond breaking in zeolites \cite{Morris2015,Roth2013} -- limits connectivity and tends to suppress large-scale organization. Vacancies -- missing links in real materials -- often compromise emergent phases by disrupting connectivity and collective behavior \cite{Simonov2020}. Such defects are known to modify electronic, magnetic, and structural properties in oxides, alloys, and skyrmion crystals \cite{Chen2016, Sun2021, Zhou2023}. Despite their relevance, understanding whether shortcuts and sparsity lead to structural transitions in complex systems remains a crucial open challenge.

In the context of complex networks, shortcuts were first formalized in the celebrated Watts-Strogatz (WS) model \cite{Watts1998}. There, a simple ring lattice is ‘rewired’ to introduce increasing amounts of disorder, allowing for a middle ground between regular and random networks while preserving high clustering and a very slowly growing diameter \cite{Newman1999,Newman2000}. Extensions of the WS model, where the rewiring probability decays with distance, have further demonstrated that such shortcuts can modify the spectral dimension of the underlying graph \cite{Millan2021}. However, do small-world effects merely induce smooth crossovers towards randomness or possibly lead to genuine structural phase transitions?


Dilute lattice models have long suggested that structural perturbations may influence universality classes \cite{Vojta2006}. Griffiths theoretically demonstrated that disordered systems could exhibit non-analytic free energy over an extended temperature range \cite{Griffiths1969}. Understanding Griffiths phases is essential for comprehending many other disordered systems, such as dilute ferroelectric materials \cite{Westphal1992,Timonin1997}. However, no general framework has emerged to predict when and how such quenched disorder induces non-ergodic behavior in arbitrary network architectures.

This motivated recent work on intrinsically scale-invariant networks, where structural fixed points were defined through the Laplacian Renormalization Group (LRG) \cite{LRG}. This framework has introduced the notion of non-integer dimension tied to a constant entropy-loss rate across scales \cite{Poggialini2025}. Crucially, this enables a distinction between scale-free and scale-invariant networks, with the latter admitting structural universality classes characterized by a well-defined 
\emph{spectral dimension}, \(d_s\). This global property of networks is related, for example, to the infrared singularity of the Gaussian process \cite{Cassi1996}, and provides a robust generalization of the standard concept of Euclidean dimension for heterogeneous systems \cite{Cassi1992,Cassi1996}. Furthermore, structural scale invariance leads to the emergence of characteristic strange attractors in reciprocal space, allowing the identification of hidden symmetries and local dimensionalities undetectable \new{via standard methods} \cite{Villegas2025}. \new{Here, we exploit the LRG to address a fundamental question: Is it possible to define attraction basins for lattices and scale-invariant networks by exploring their response to topological perturbations?}



We show that scale-invariant architectures under random perturbations (shortcut addition or link removal) exhibit previously unidentified structural phase transitions, which depend on the tiling patterns of the underlying structure. We term this phenomenon {\em geometric criticality}, as this new topological phase transition corresponds to a geometric breakdown point where the system loses its well-defined dimension. This mechanism gives rise to highly heterogeneous, lacunar fractal networks and naturally defines an attraction basin for the LRG fixed points. Our framework provides a novel lens for analyzing the interplay between quenched disorder and dynamics, enabling a systematic exploration of how different perturbations alter the dimensional properties of the system itself.

\paragraph*{\textbf{Structural fixed points.}} Topological scale-invariant structures were defined in~\cite{Poggialini2025}, based on the LRG~\cite{LRG,Gabrielli2025}, which analyzes the spectral properties of the network Laplacian \( \hat{L} = \hat{D} - \hat{A} \), where \( \hat{A} \) is the adjacency matrix and \( \hat{D} \) is the diagonal matrix \new{of node degrees}. Using the diffusion operator \( e^{-\tau \hat{L}} \) as the evolution of a heat kernel at scale/diffusion time \( \tau \), one defines the Laplacian density matrix~\cite{Domenico2016} $\hat{\rho}(\tau) = \frac{e^{-\tau \hat{L}}}{Z(\tau)}$, with $Z(\tau) = \mathrm{Tr}\left[e^{-\tau \hat{L}}\right],$ playing the role of a partition function~\cite{InfoCore}.
This allows for a statistical mechanical description of the network structure, in which \( \tau \) plays the role of inverse temperature and \( \hat{L} \) acts as a Hamiltonian. The associated entropy $S(\tau) = -\mathrm{Tr}[\hat{\rho}(\tau)\log \hat{\rho}(\tau)]$ quantifies the effective number of active degrees of freedom at a given scale \( \tau \), decreasing from $S=\log N$ at $\tau=0$ to $S=0$ at $\tau\rightarrow\infty$. Its derivative with respect to \( \log \tau \) defines the entropic susceptibility or heat capacity, $C(\tau) = -\frac{dS}{d \log \tau}$, which measures the rate of information loss as the system is coarse-grained~\cite{InfoCore,LRG}.

\begin{figure}[hbtp]
    \centering
    \includegraphics[width=1.0\linewidth]{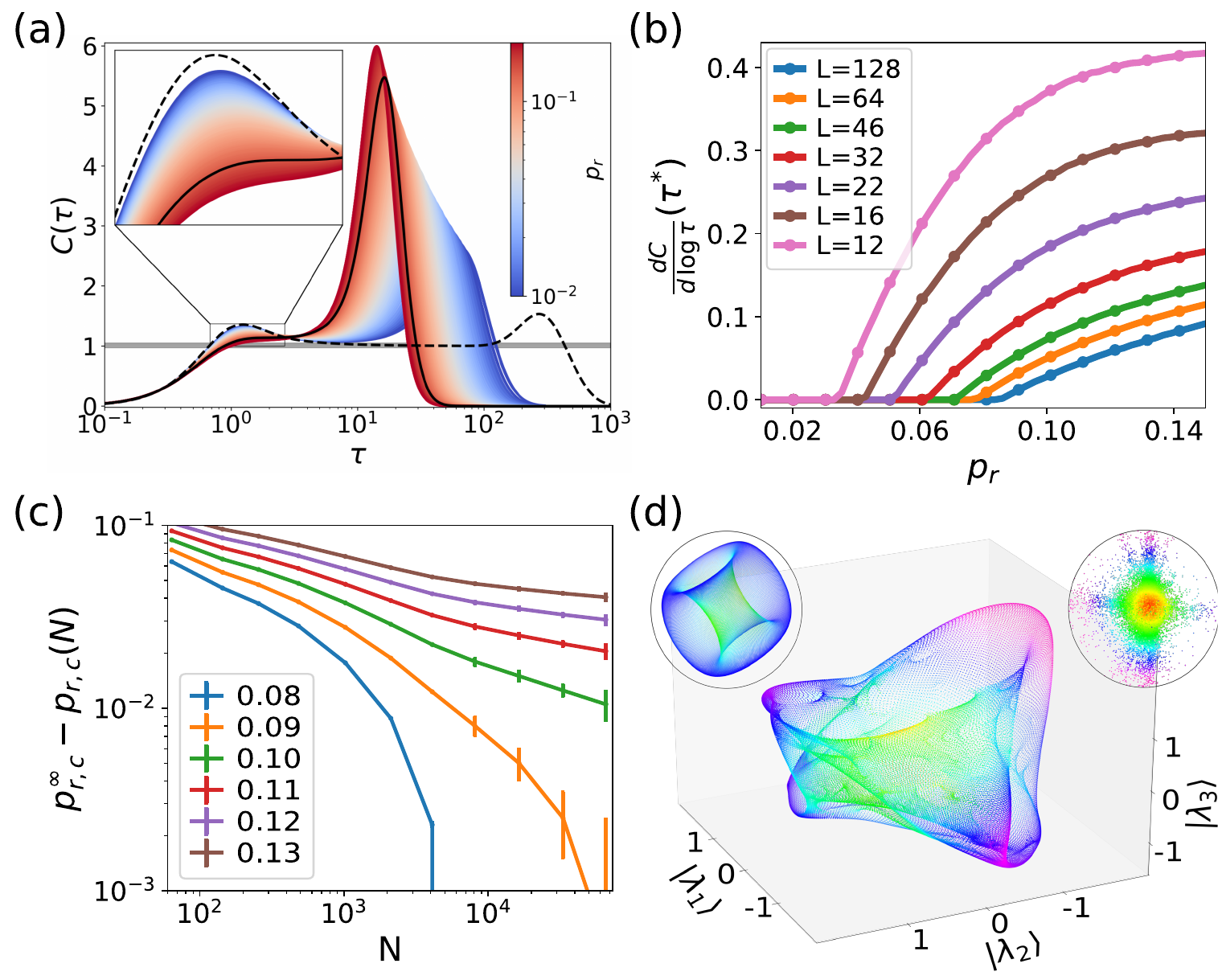}
    \caption{Shortcuts in 2D square lattices.
    \textbf{(a)} Heat capacity, $C$, versus diffusion time, $\tau$, for different rewiring probabilities (see colorbar, $L=64$). The black dashed line corresponds to the unperturbed lattice. Inset: Zoom of the $\Lambda$-region. \textbf{(b)} First derivative of the heat capacity at the first peak versus rewiring probability, $p_r$, for different lattice sizes (see legend). \textbf{(c)} Finite-size scaling analysis of the vanishing peaks in (b). The distance of the size-dependent peak locations $p_{r,c}(N)$ from their asymptotic value for $N\rightarrow\infty$, $p_{r,c}^\infty$, scales as a power law of the system size, only at $p_{r,c}^\infty=0.10(1)$, revealing the existence of true scaling at criticality. \textbf{(d)} Low-dimensional representation using the three smallest low-frequency normalized network modes as the coordinate axes for a 2D square lattice of size $L=256$. Parameters: $p_{r}=2.3\times10^{-4}$, $p_{r}=0$ (left inset), and $p_{r}=0.1$ (right inset). All curves have been averaged over $10^{3}$--$10^{5}$ realizations.}
    \label{SW-nets}
\end{figure}

Scale invariance is defined by the condition that \( C(\tau) \) remains constant across a broad range of scales, indicating the absence of a characteristic structural scale in the system~\cite{Poggialini2025}. \new{This implies an entropy-loss rate proportional to} the spectral dimension \( d_s \), a global geometric quantity that generalizes the notion of Euclidean dimension to networks~\cite{Cassi1992,Cassi1996}. Hence, the presence of a plateau in the heat capacity,  $C_0=d_s/2$, reveals the spectral dimension $d_s$ that characterizes the scale-invariant nature of homogeneous lattices or networks. Moreover, $C$ usually presents small-scale resonant modes at short times, a signature of the elementary small-scale structure of the lattice or network~\cite{TBM}, i.e., the ultraviolet cut-off $\Lambda$ (in the jargon of the Renormalization Group). \new{For a 2D square lattice, this corresponds to the peak of $C$ at $\tau \approx 1.24$ (black dashed line in Fig.~\ref{SW-nets}(a)).}

The second peak at larger values of $\tau$ instead reflects the whole lattice scale. In particular, the recent categorization of structural fixed points has shed light on four main classes of scale-invariant networks \cite{Poggialini2025}: (i) regular lattices, (ii) connected loopless networks, i.e., trees \cite{Donetti2004}, (iii) those with a non-vanishing clustering coefficient, as (u,v) flowers \cite{Rozenfeld2007} (including the Dorogovtsev-Goltsev-Mendes network \cite{Dorogovtsev2002} (DGM)), and Kim and Holme networks \cite{KimHolme} and, (iv) hierarchical modular networks (HMNs) originally proposed taking inspiration from brain networks \cite{Moretti2013}.

The ordered set of Laplacian eigenvalues provides further geometric insight, as it represents network frequency modes analogous to the usual Fourier analysis~\cite{LRG,Villegas2025}. Each eigenvalue represents a characteristic oscillation scale: low eigenvalues correspond to large-scale (low-frequency) modes, while high ones capture small-scale fluctuations (high frequencies). The spectral dimension inferred from entropy loss governs Laplacian embeddings, allowing any network to be unfolded into a low-dimensional manifold in \(\kappa\)-space. Using the first $n$ nontrivial Laplacian eigenvectors (for any connected network, the smallest eigenvalue, \(\lambda_0 = 0\), is uniform and does not carry structural information), it is possible to \new{map network nodes into \(\mathbb{R}^n\), providing a geometric representation}. In these embeddings, scale-invariant networks display attractor-like structures where the correlation dimension $D$ matches the Euclidean dimension $d_E$ for lattices, and $D = d_S/2$ for heterogeneous networks \cite{Villegas2025}. This provides a minimal and robust geometric representation \new{just considering the minimum integer embedding dimension $d_E \ge D$} \cite{Villegas2025}. 

\paragraph*{\textbf{Topological shortcuts.}}
We first consider the general case of regular lattices, where long-range connections are introduced by rewiring a fraction of the links. \new{This is achieved by independently rewiring each link endpoint with probability $p_r$}, while preserving the total number of edges in the network. Fig.~\ref{SW-nets}(a) shows the heat capacity as a function of the diffusion time for a 2D square network with varying rewiring probabilities. From now on, we consider the resulting giant component of the perturbed structure. As the rewiring probability increases, the plateau in the heat capacity narrows, until the early-time heat capacity peak vanishes, \new{signaling the breakdown of translational symmetry} (see the saddle point on the solid black curve in the inset).
Considering this first peak as a hallmark of lattice-scale order, we quantify this transition by examining the derivative of the heat capacity at the characteristic time of the first peak (see Fig.~\ref{SW-nets}(b)). Through finite-size scaling analyses, as reported in Fig.~\ref{SW-nets}(c), we confirm the existence of a finite critical rewiring probability in the thermodynamic limit, $p_{r,c}^{sq}=0.10(1)$, which naturally defines the attraction basin of the 2D square lattice under the introduction of shortcuts. This geometric deformation can be further visualized using the 3D Laplacian projection introduced above. As reported in Fig.~\ref{SW-nets}(d), the 2D lattice becomes distorted due to long-range connections, leading to a collapsed Gaussian-like network at the critical value (see insets of Fig.~\ref{SW-nets}(d) and Supplementary Videos \cite{SM}). A similar behavior is observed for other regular lattices, such as 2D triangular and hexagonal lattices (see Appendix \ref{2DApp}), although with different critical rewiring thresholds: $p_{r,c}^{tr}=0.17(4)$ and $p_{r,c}^{hex}=0.055(10)$, respectively.
Taken together, the disappearance of the short-scale peak in the heat capacity and the Gaussian collapse observed in Laplacian space constitute complementary signatures of a structural phase transition. These jointly define the limits of the attraction basin of the LRG structural fixed points.

\begin{figure}[hbtp]
    \centering
    \includegraphics[width=1.0\linewidth]{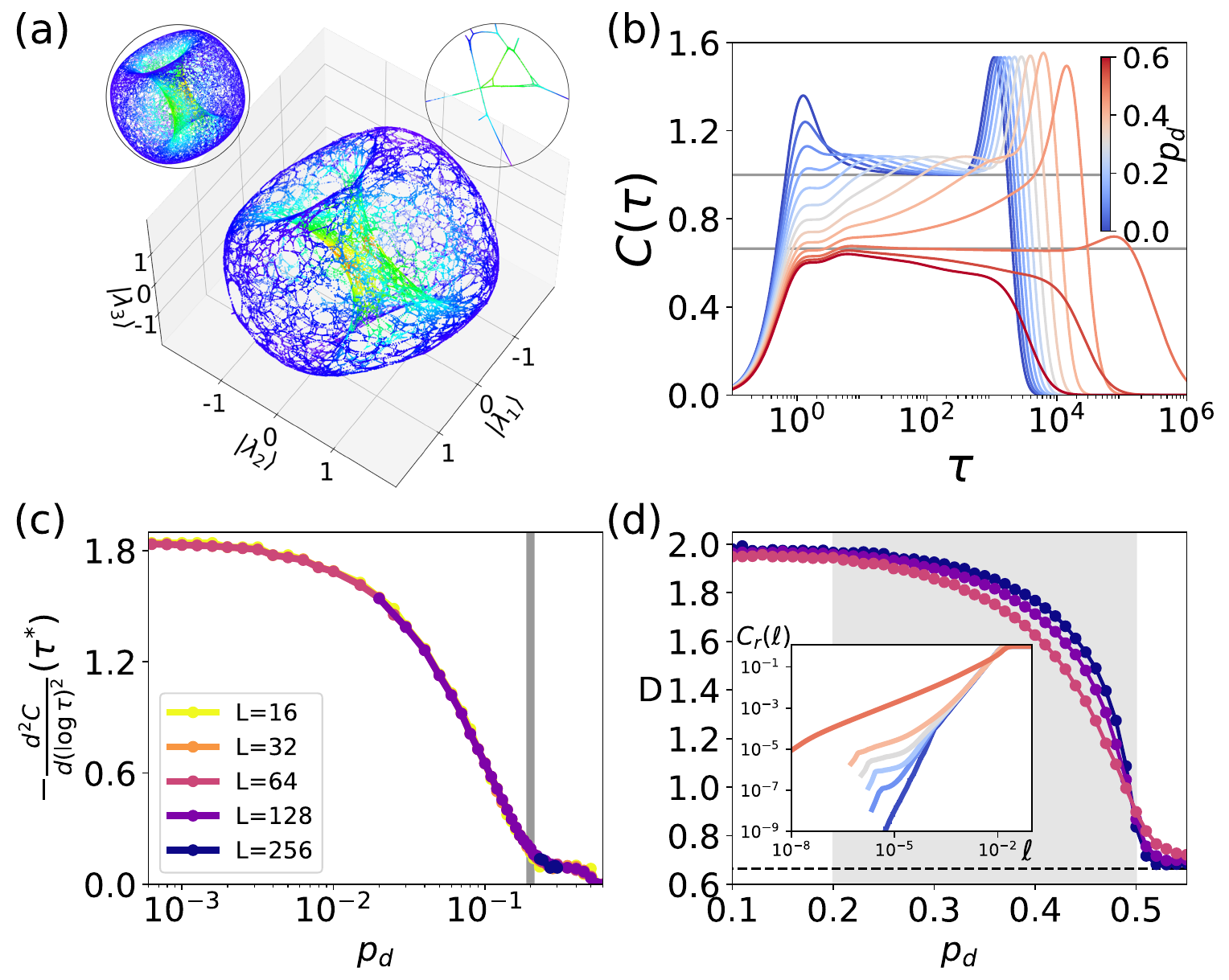}
    \caption{Diluted 2D square lattices. \textbf{(a)} Low-dimensional representation using the three smallest low-frequency normalized network modes as the coordinate axes, for a lattice with $p_{d}=0.35$, $p_{d}=0.2$ (left inset), and $p_{d}=0.5$ (right inset). \textbf{(b)} Heat capacity, $C$, versus diffusion time, $\tau$, for different dilute probabilities (see colorbar). Gray lines represent the expected plateau for a 2D lattice and a RT.  \textbf{(c)} Second derivative of the heat capacity at the first peak versus dilute probability, $p_d$, for different lattice sizes (see legend). The vertical gray line represents the point where the 2D ultraviolet cut-off is expected to disappear. \textbf{(d)} Estimated correlation dimension, $D$, versus dilution probability, $p_d$ for different lattice sizes. Inset: Correlation integral $C(\ell)$ vs distance $\ell$ for different dilute probabilities. The gray area represents dilute probabilities between $p_{d,c}$ and the percolation threshold. All curves have been averaged over $10^{3}$--$10^{4}$ realizations.}
    \label{Dilute-models}
\end{figure}

\paragraph*{\textbf{Structural sparsity.}}
We now turn to the role of link dilution, a classical model for quenched disorder known to influence critical behavior in various universality classes \cite{Vojta2006}. Starting from a regular lattice, we remove links randomly with probability $p_d$. Fig.~\ref{Dilute-models}(a) displays the Laplacian embedding of a 2D square lattice at different values of $p_d$ up to the percolation threshold, $p_c$, where the giant connected component disappears ($p_c=0.5$ for the 2D square lattice), and a random tree (RT) structure emerges. This is a direct consequence of the Alexander-Orbach conjecture \cite{AO}, which predicts that, at criticality, the giant component for isotropic percolation has a universal spectral dimension $d_s=\nicefrac{4}{3}$ across all dimensions.

The analysis of heat capacity versus dilution probability unveils a specific transition point, $p_{d,c}$, at which the ultraviolet cut-off $\Lambda$ vanishes. Beyond this point, the effective dimension of the network decreases progressively toward the RT value $ d_s = 4/3$ at $p_c$. We detect this geometric breakdown by analyzing the curvature (concavity) of the first peak in the heat capacity (see Fig.~\ref{Dilute-models}(b)), as illustrated in Fig.~\ref{Dilute-models}(c).

To better characterize this phase transition, we also compute the correlation integral as defined by Grassberger and Procaccia~\cite{CorrDim,ScaleFree,Villegas2025}. 
\new{Using the projection of the $i-{th}$}  node onto the $d$-dimensional space spanned by the first $d$ non-trivial Laplacian eigenvectors, \(\mathbf{x}_i = \sqrt{N}\left( \langle \lambda_1 | i \rangle, \langle \lambda_2 | i \rangle, \ldots, \langle \lambda_d | i \rangle \right)\), we can define the correlation integral in the reciprocal space as \new{$C(\ell) = \frac{2}{N(N-1)} \sum_{i>j} \Theta(\ell - \| \mathbf{x}_i - \mathbf{x}_j \|) \sim \ell^D$}, where $\Theta$ is the Heaviside step function.
If the scaling law $C(\ell) \propto \ell^D$ holds, then $D$ defines the network correlation dimension, which enables the detection of significant topological alterations in the perturbed network.
As shown in Fig.~\ref{Dilute-models}(d), $D$ remains stable up to $p_{d,c}$, i.e., for low dilution, the lattice does not lose its bidimensional structure at large scale. From that point on, the dimensionality progressively decreases until it reaches the expected dimension for a RT.
Both analyses capture the same phenomenon, yielding compatible estimates for the critical dilution probability $p_{d,c}=0.20(5)$ for the 2D square lattice.
Again, an equivalent behavior emerges for triangular and hexagonal lattices (see Appendix \ref{2DApp}), but with different critical points: $p_{d,c}^{tr}=0.25(1)$ and $p_{d_c}^{hex}=0.11(2)$, respectively. \new{For 3D cubic lattices, we estimate a critical dilution threshold of $p_{d,c}=0.55(5)$, while the rewiring threshold was found to be $p_{r,c}=0.20(5)$ (see Appendix \ref{3DApp}).} Note that performing correlation dimension analyses in the presence of shortcuts is considerably more challenging. While the method remains valid, the scaling region contracts, statistical noise increases, and extensive averaging becomes necessary. Moreover, the loss of spatial locality impairs the accuracy of Laplacian embeddings, making it more difficult to identify an effective embedding dimension. Conversely, in dilute models, it is another complementary measure that allows for the analysis of increasing system sizes, as it depends only on a few eigenvectors. 

\paragraph*{\textbf{Heterogeneous Networks.}}
We now extend our analysis to heterogeneous structures. \new{For} generic trees, the critical rewiring threshold scales as \( p_{r,c} \sim \mathcal{O}(\nicefrac{1}{N}) \), while link deletion leads to a trivial threshold \( p_{d,c} = 1 \), since the spectral dimension remains unchanged (see Appendix \ref{RTApp}). In contrast, DGM networks exhibit more significant and non-trivial behavior. \new{For these, we identify non-trivial critical thresholds for rewiring and dilution:} $p_{r,c}\neq0$ (see Appendix \ref{DGMApp} for a discussion on the issue) and $p_{d,c}=0.15(5)$, respectively (the percolation threshold is $p_c=0.95(5)$, see Appendix \ref{DGMApp}). Figs.~\ref{Het-nets}(a) and (c) show the correlation dimension, $D$, as a function of $p_d$ for DGM and HMN networks. In DGM networks, $D$ remains constant up to $p_{d,c}$ and then begins to decrease. In particular, these networks show evidence of flowing to the vicinity of an unstable structural fixed point for $p_d=0.75(5)$: as shown in Fig.~\ref{Het-nets}(b), their structure converges to that of a Barab\'asi--Albert (BA) network, following a power-law degree distribution $P(\kappa)\sim\kappa^{-3}$. Furthermore, the heat capacity stabilizes at $C_0=1$, matching the BA spectral signature \cite{LRG,Poggialini2025} (see Fig.~\ref{Het-nets}(b)). In HMNs (see Appendix \ref{HMNApp} for further analysis), a similar decay in $D$ is observed beyond the corresponding $p_{d,c}$. However, unlike DGM networks, after the percolation threshold, $D$ stabilizes at the RT spectral dimension, $d_{s}=4/3$. The absence of hubs prevents the system from flowing close to the BA unstable fixed point reported in the DGM case.

\begin{figure}[hbtp]
    \centering
    \includegraphics[width=1.0\linewidth]{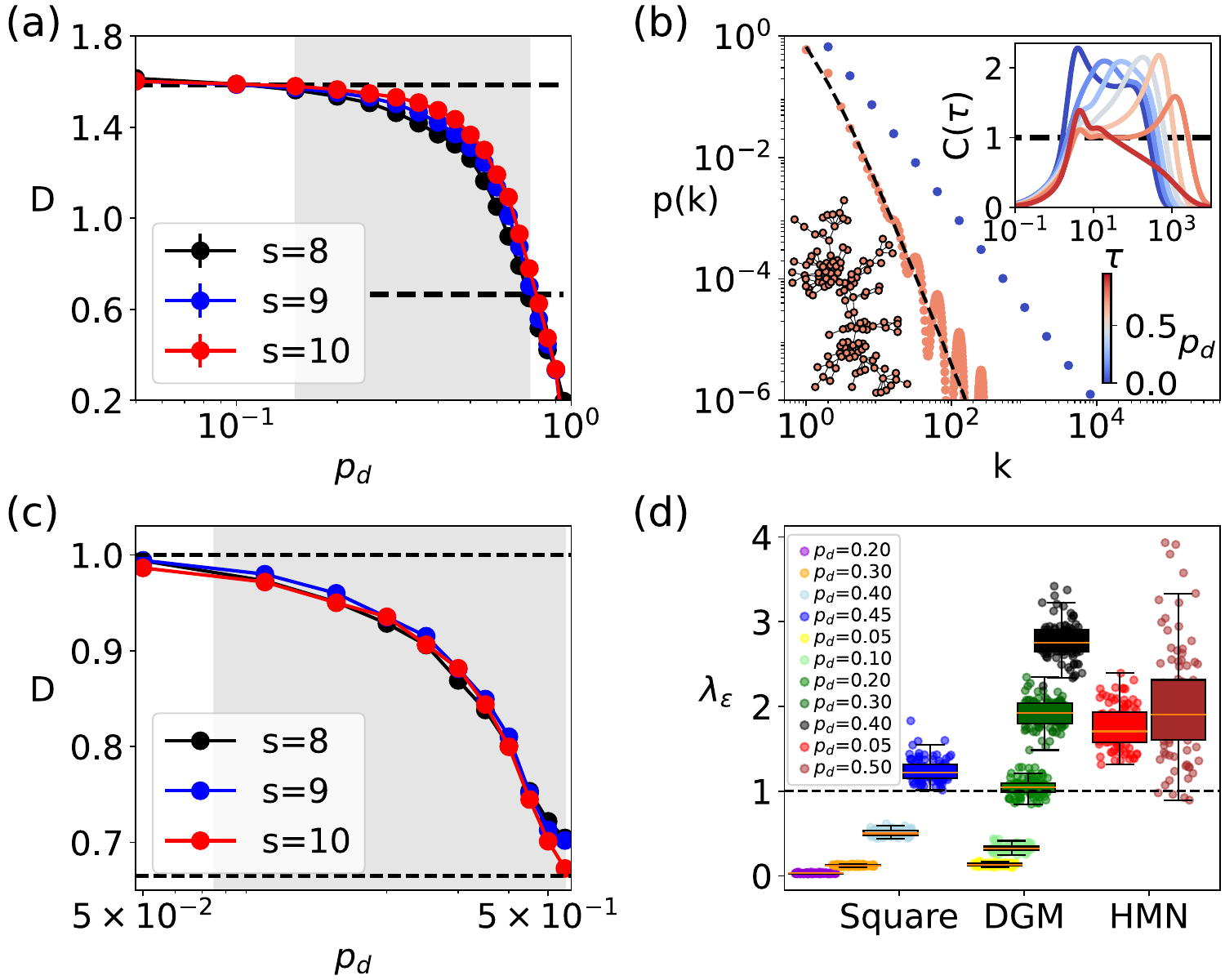}
    \caption{Diluted scale-invariant networks.
    \textbf{(a, c)} Estimated correlation dimension, $D$, versus dilution probability, $p_d$, for networks of different generation levels, $s$ (see legend), for: \textbf{(a)} DGM ($N_s=\frac{3+3^s}{2}$) and \textbf{(c)} HMN with $m_0=3$ and $\alpha=4$ ($N_s=2^s m_0$).
    \textbf{(b)} Degree distribution, $P(\kappa)$, versus node degree, $\kappa$, for a DGM network with $p_d=0$ (blue points) and $p_d=0.75$ (orange points). The black dashed line indicates the power-law scaling of a BA network, $P(\kappa)\propto\kappa^{-3}$. Inset: Specific heat for different dilution probabilities; for $p_d=0.75$, the plateau at $C_0=1$ confirms the BA limit.
    \textbf{(d)} \new{Lacunarity index, $\lambda$, calculated at a fixed scale $\epsilon=10^{-2}$ for different network topologies (x-axis). Colors distinguish different dilution probabilities (see legend). The horizontal dashed line marks the theoretical reference $\lambda=1$ (where variance equals mean).}
    All curves have been averaged over $10^{3}$--$10^{4}$ realizations.}
    \label{Het-nets}
\end{figure}
Finally, to further characterize the loss of structural regularity, we compute the lacunarity \cite{Mandelbrot, GabrielliBook}, $\lambda_\varepsilon$, or void distribution, defined as the square coefficient of variation of pixels per box in a box-counting analysis: $\lambda_\varepsilon=\frac{\sigma _{\varepsilon }^{2}}{\mu _{\varepsilon }^{2}}$, with $\varepsilon$ the box-size.
As seen in Fig.~\ref{Het-nets}(d), by using a 3D embedding space, lacunarity increases with dilution across all studied networks. This indicates a breakdown of statistical self-similarity and the emergence of heterogeneity in the spatial complexity. For HMNs, high lacunarity is present even without dilution, indicating their intrinsic geometric irregularity. This is particularly significant because when lacunarity is low, fractals behave similarly to analytically continued hypercubic lattices~\cite{Gefen1983}. Therefore, lacunarity may serve as a signature of anomalous universal behavior.

\paragraph*{\textbf{Outlook.}} It is generally understood in critical phenomena that once a symmetry of the order parameter and the interaction range are fixed, only the dimension controls the critical properties of the system \cite{Binney,Amit,Kadanoff1971}. In the framework of complex networks, this perspective has recently been extended to include scale-invariant networks defined by a well-defined spectral dimension \cite{Poggialini2025}. 
However, it is unclear how topological noise affects dimensionality in such systems. 

Here, we uncover a novel class of structural phase
transitions,  which arise in scale-invariant networks and lattices undergoing a critical geometric breakdown under topological perturbations. This allows us to introduce the concept of geometric criticality, \new{occurring} when lattices lose translational invariance and networks lose the homogeneous spectral structure
We further demonstrate the existence of attraction basins that depend not only on dimension but also on microscopic structural details, such as the lattice tiling. Topological shortcuts act as local perturbations that induce a global collapse of the underlying manifold once a critical threshold is crossed. Conversely, link dilution erodes dimensional uniformity, leading to multifractal-like geometries. For link dilution, this breakdown also generates high lacunarity values, a measure originally proposed to analyze the deviation of a fractal from being translationally invariant \cite{Gefen1980}, which has been shown to impact critical phenomena on fractal lattices \cite{Gefen1983}. 

For HMNs, which are known to exhibit Griffiths phases due to their hierarchical heterogeneity~\cite{Moretti2013}, we provide the first evidence of scale-invariant networks that inherently possess high lacunarity, even in the absence of external perturbations. This supports earlier arguments that critical phenomena in fractal systems depend not only on dimension but also on more intricate geometric and topological features. In particular, our framework opens the door to an extensive analysis of Griffiths phases and broad-critical regions from a rigorous perspective.

These geometric transitions, although purely structural in origin, are likely a signature of strong dynamical consequences, potentially altering well-known processes such as transport, synchronization, or memory properties in real systems. We hope that this work will stimulate future research in this direction.
\newpage

\begin{acknowledgments}
\paragraph*{\textbf{Acknowledgments.}} We thank M.A. Mu\~noz and A. Gabrielli for useful discussions and comments. P.V. acknowledges the Spanish Ministry of Research and Innovation and Agencia Estatal de Investigación (AEI), MICIN/AEI/10.13039/501100011033, for financial support through Project PID2023-149174NB-I00, funded also by European Regional Development Funds, and  Ref. PID2020-113681GB-I00.
\end{acknowledgments}

\appendix
\renewcommand{\appendixname}{APPENDIX}
\section{Geometric criticality in different networks and lattices}
Table~\ref{tab:percolation_thresholds} reports the critical values marking the onset of geometric criticality across diverse lattice and network topologies. We examine structural perturbations driven by dilution and rewiring (at conserved link density). These estimates derive from a joint evaluation of the network heat capacity and the correlation dimension, within the limits of the latter's interpretability as detailed in the main text.

\begin{widetext}
\vspace{-.5cm}
\begin{center}
\begin{table}[H]
\centering
\begin{tabular}{ccccccc}
\hline 
 & 2D square  & 2D triangular  & 2D hexagonal  & 3D cubic & Trees & DGM\tabularnewline
\hline 
\hline 
Rewiring & 0.10(1)  & 0.17(4)  & 0.055(10)  & 0.20(5) & $\mathcal{O}(\nicefrac{1}{N})$ & 0.05(3)\tabularnewline
\hline 
Dilution & 0.20(5) & 0.25(1) & 0.11(2) & 0.55(5) & 1 & 0.15(5)\tabularnewline
\hline 
\end{tabular}
\caption{Geometric criticality thresholds for structural perturbations. The table reports the critical values for rewiring and dilution processes across various regular lattices and complex network topologies. Parentheses indicate the estimated uncertainty.}
\vspace{-.8cm}
\label{tab:percolation_thresholds} 
\end{table}
\end{center}
\end{widetext}

\section{2D lattices}
\label{2DApp}
\subsection{Triangular lattice}

\begin{figure}[hbtp]
    \centering
    \includegraphics[width=1.0\linewidth]{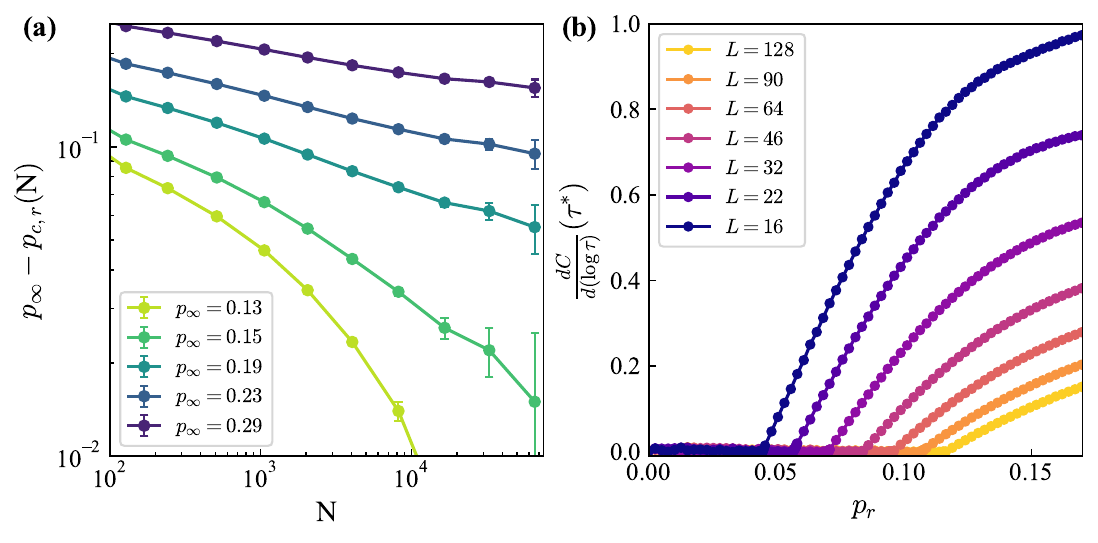}
    \caption{Criticality in 2D triangular lattices under rewiring.
\textbf{(a)} Finite-size scaling of the heat capacity peak locations. The convergence of the size-dependent critical points $p_{r,c}(N)$ toward the asymptotic limit $p_{r,c}^\infty$ follows a power law, uniquely identifying the critical threshold at $p_{r,c} = 0.17(4)$ in the thermodynamic limit.
\textbf{(b)} First derivative of the heat capacity with respect to the rewiring probability $p_r$, shown for different lattice sizes (see legend).}
    \label{2DTriRew}
\end{figure}

We present a detailed analysis of 2D triangular lattices here. Fig.~\ref{2DTriRew} reports finite-size scaling analyses confirming the existence of a finite critical rewiring probability in the thermodynamic limit, $p_{r,c}=0.17(4)$. As in the case of the square lattice, this threshold delimits the attraction basin of the 2D triangular lattice under the introduction of shortcuts. Note that the specific value of $p_{r,c}$ depends on the underlying lattice tiling.

\begin{figure}[H]
    \centering
    \includegraphics[width=0.85\linewidth]{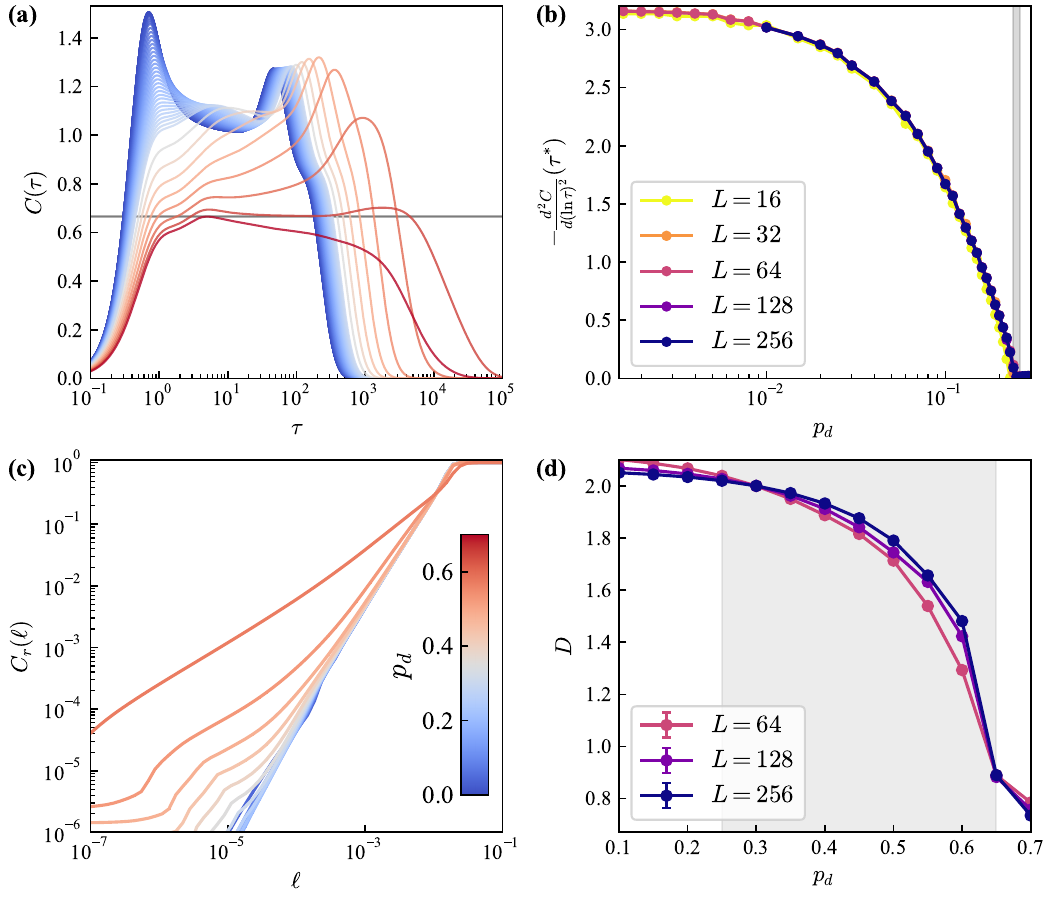}
    \caption{Geometric breakdown of 2D triangular lattices under dilution.
\textbf{(a)} Heat capacity $C$ as a function of diffusion time $\tau$ for varying dilution probabilities (see colorbar). The gray line marks the Alexander-Orbach (RT) plateau.
\textbf{(b)} Second derivative (curvature) of the heat capacity at the first peak versus dilution probability $p_d$ for different lattice sizes. The vertical gray line indicates the critical point where the 2D ultraviolet cut-off vanishes.
\textbf{(c)} Correlation integral $C_{r}(\ell)$ versus distance $\ell$ for different dilution probabilities.
\textbf{(d)} Estimated correlation dimension $D$ versus $p_d$ for different lattice sizes. The gray shaded region marks the regime between $p_{d,c}$ and the percolation threshold. All curves represent averages over $10^{3}$--$10^{4}$ realizations.}
    \label{CSDilTri}
\end{figure}

In the case of 2D triangular lattices, the heat capacity profile reveals the specific transition point $p_{d,c}$ marking the collapse of the ultraviolet cut-off $\Lambda$. This geometric breakdown is signaled by the curvature (concavity) of the first heat capacity peak, illustrated in Figs.~\ref{CSDilTri}(a) and (b). Regarding the network topology, Fig.~\ref{CSDilTri}(c) reports the correlation dimension as a function of rewiring probability, with the specific scaling exponent $C(\ell) \propto \ell^D$ shown in Fig.~\ref{CSDilTri}(d). Crucially, for dilutions exceeding $p_{d,c}$, the effective dimension progressively converges to the Alexander-Orbach (or Random Tree) value $d_s = 4/3$ at the percolation threshold $p_c$.

\subsection{Hexagonal lattice}

\begin{figure}[hbtp]
    \centering
    \includegraphics[width=1.0\linewidth]{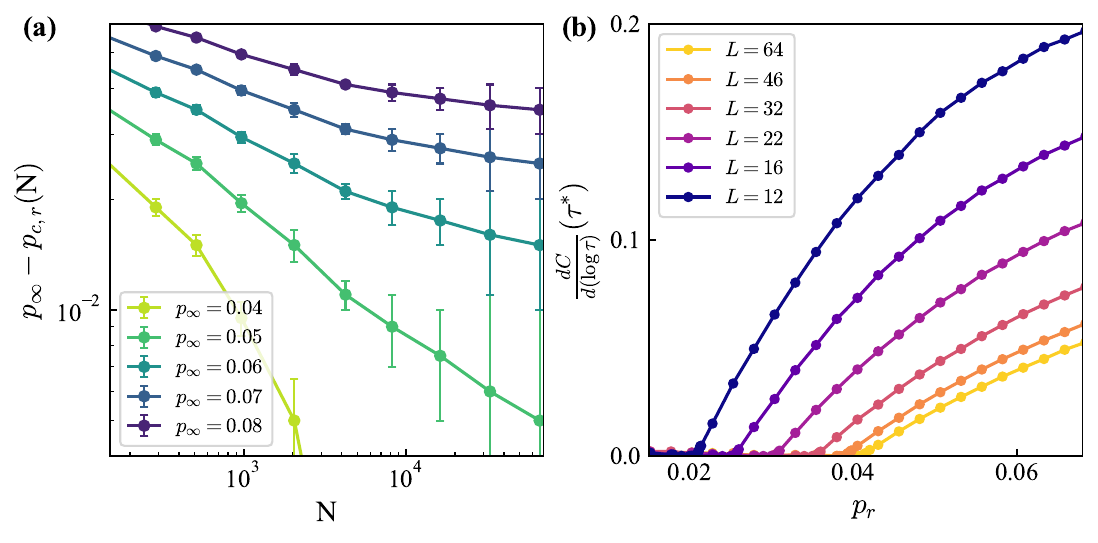}
    \caption{Criticality in 2D hexagonal lattices under rewiring.
\textbf{(a)} Finite-size scaling of the heat capacity peak locations. The convergence of the size-dependent critical points $p_{r,c}(N)$ toward the asymptotic limit $p_{r,c}^\infty$ follows a power law, uniquely identifying the critical threshold at $p_{r,c}^\infty=0.055(10)$.
\textbf{(b)} First derivative of the heat capacity with respect to the rewiring probability $p_r$, shown for different lattice sizes (see legend).}
    \label{2DHexRew}
\end{figure}

Turning to the 2D hexagonal lattice, the finite-size scaling analysis in Fig.~\ref{2DHexRew} confirms the existence of a finite critical rewiring probability in the thermodynamic limit, $p_{r,c}=0.055(10)$. As observed for square and triangular lattices, this threshold delimits the attraction basin of the hexagonal geometry against the introduction of shortcuts. These results further illustrate that the specific value of $p_{r,c}$ is intrinsically dependent on the lattice tiling.

\begin{figure}[hbtp]
    \centering
    \includegraphics[width=0.9\linewidth]{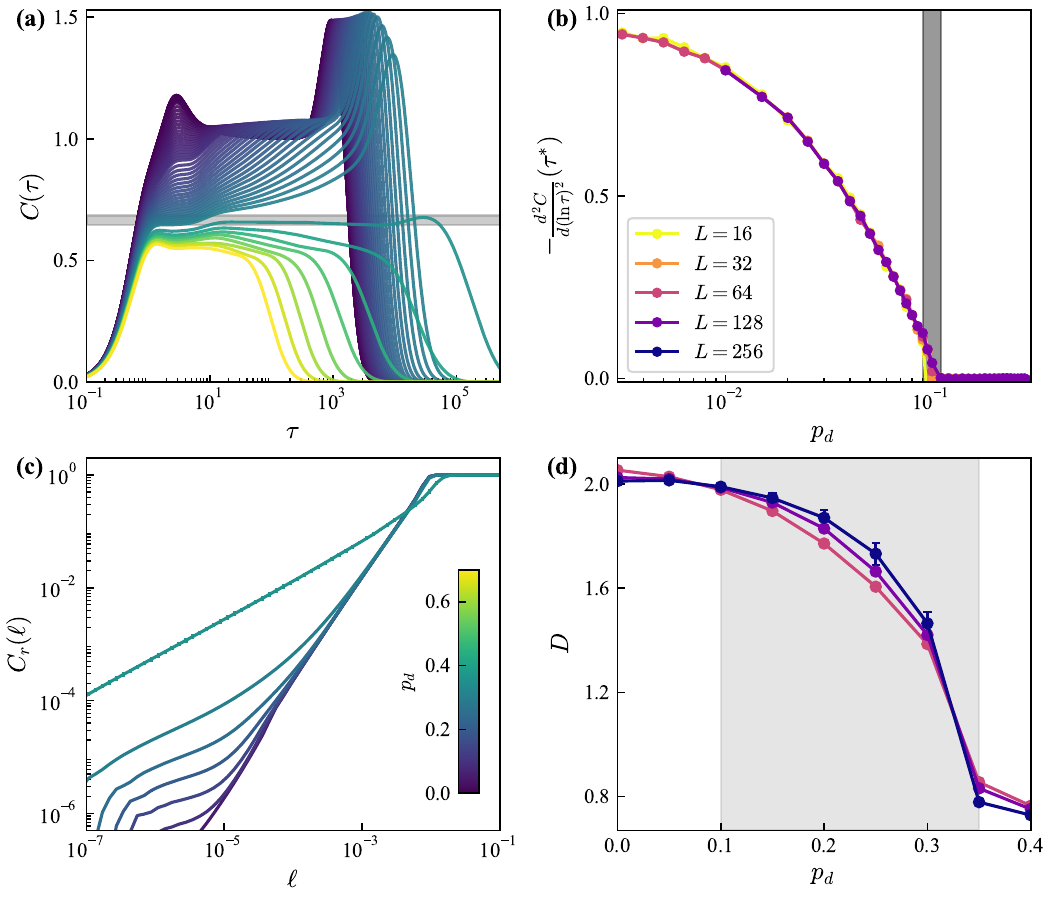}
    \caption{Geometric breakdown of 2D hexagonal lattices under dilution.
\textbf{(a)} Heat capacity $C$ as a function of diffusion time $\tau$ for varying dilution probabilities (see colorbar). The gray line marks the Alexander-Orbach (RT) plateau.
\textbf{(b)} Second derivative (curvature) of the heat capacity at the first peak versus dilution probability $p_d$ for different lattice sizes. The vertical gray line indicates the critical point where the 2D ultraviolet cut-off vanishes.
\textbf{(c)} Correlation integral $C_{r}(\ell)$ versus distance $\ell$ for different dilution probabilities.
\textbf{(d)} Estimated correlation dimension $D$ versus $p_d$ for different lattice sizes. The gray shaded region marks the regime between $p_{d,c}$ and the percolation threshold. All curves represent averages over $10^{3}$--$10^{4}$ realizations.}
    \label{CSDilHex}
\end{figure}

For 2D hexagonal lattices, the analysis of the heat capacity as a function of the dilution probability identifies the critical point $p_{d,c}$ at which the ultraviolet cut-off $\Lambda$ vanishes. We detect this geometric breakdown by analyzing the concavity of the first heat capacity peak (see Fig.~\ref{CSDilHex}(a)), as detailed in Fig.~\ref{CSDilHex}(b). Figure~\ref{CSDilHex}(c) presents the correlation dimension under rewiring, while Fig.~\ref{CSDilHex}(d) plots the dimension derived from the scaling $C(\ell) \propto \ell^D$. Notably, beyond $p_{d,c}$, the effective dimension decreases progressively toward the random tree (RT) value $d_s = 4/3$ at $p_c$.

\section{3D lattices}
\label{3DApp}

\begin{figure}[hbtp]
    \centering
    \includegraphics[width=0.95\linewidth]{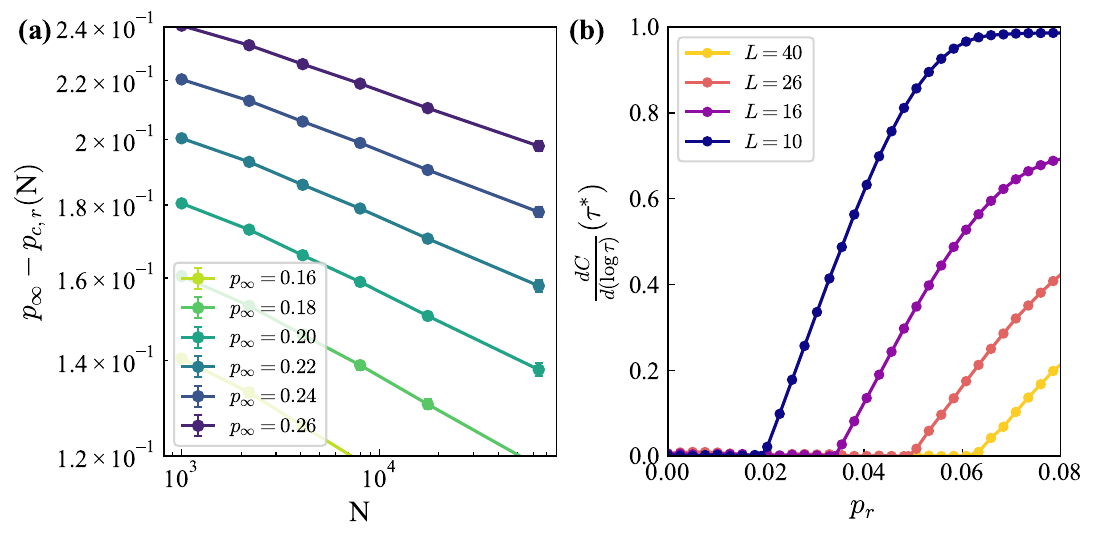}
    \caption{Criticality in 3D cubic lattices under rewiring.
\textbf{(a)} Finite-size scaling of the heat capacity peak locations. The convergence of the size-dependent critical points $p_{r,c}(N)$ toward the asymptotic limit $p_{r,c}^\infty$ follows a power law, uniquely identifying the critical threshold at $p_{r,c}^\infty=0.20(5)$.
\textbf{(b)} First derivative of the heat capacity with respect to the rewiring probability $p_r$, shown for different lattice sizes (see legend).}
    \label{3DRew}
\end{figure}

Extending our analysis to three dimensions, Fig.~\ref{3DRew} illustrates the impact of shortcuts on cubic lattices. Finite-size scaling confirms the existence of a critical rewiring threshold at $p_{r,c}=0.20(5)$ in the thermodynamic limit. This value marks the boundary of the lattice's attraction basin, analogous to the 2D cases. These results reinforce the observation that the critical point is intrinsically linked to the specific lattice geometry.

\begin{figure}[hbtp]
    \centering
    \includegraphics[width=1\linewidth]{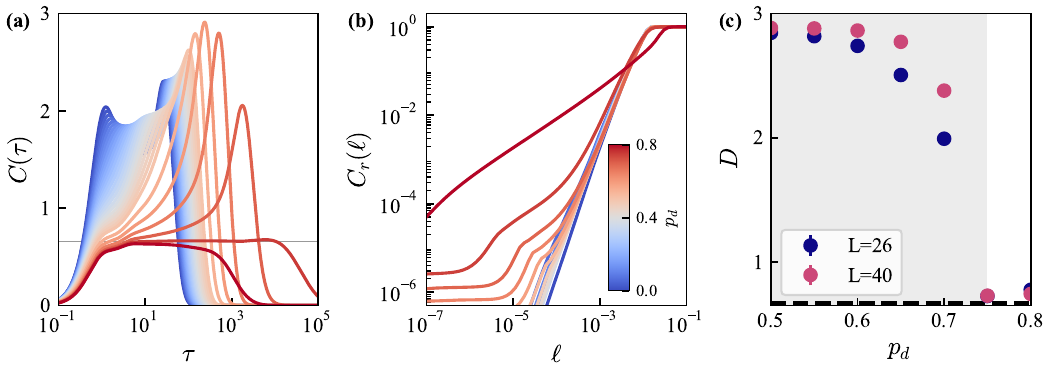}
    \caption{Geometric breakdown of 3D cubic lattices under dilution.
\textbf{(a)} Heat capacity $C$ versus diffusion time $\tau$ for varying dilution probabilities (see colorbar). The gray line marks the Alexander-Orbach (RT) plateau.
\textbf{(b)} Correlation integral $C_{r}(\ell)$ as a function of distance $\ell$ for different dilution probabilities.
\textbf{(c)} Estimated correlation dimension $D$ versus dilution probability $p_d$ for different lattice sizes. The gray shaded region marks the interval between $p_{d,c}$ and the percolation threshold. All curves represent averages over $10^{3}$--$10^{4}$ realizations.}
    \label{CSDil3D}
\end{figure}

For 3D cubic lattices, analyzing the heat capacity as a function of the dilution probability identifies the critical point $p_{d,c}$ at which the ultraviolet cut-off $\Lambda$ vanishes. We detect this geometric breakdown by examining the concavity of the first heat capacity peak, as illustrated in the left panel of Fig.~\ref{CSDil3D}. Figure~\ref{CSDil3D} also presents the correlation dimension under rewiring, while the specific dimension extracted from the scaling $C(\ell) \propto \ell^D$ is reported in the rightmost panel. Notably, beyond $p_{d,c}$, the effective dimension of the network progressively decreases toward the random tree (RT) value $d_s = 4/3$ at $p_c$.

\section{DGM networks}
\label{DGMApp}
\begin{figure}[H]
    \centering
    \includegraphics[width=1.0\linewidth]{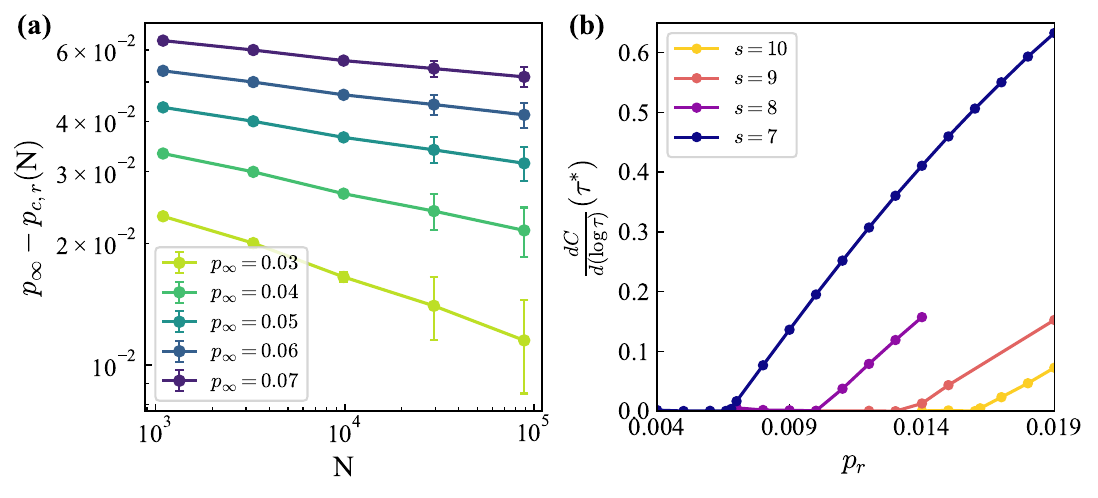}
    \caption{Criticality in DGM networks under rewiring.
\textbf{(a)} Finite-size scaling of the heat capacity peak locations. The convergence of the size-dependent critical points $p_{r,c}(N)$ toward the asymptotic limit $p_{r,c}^\infty$ follows a power law, uniquely identifying the critical threshold at $p_{r,c}^\infty=0.05(3)$.
\textbf{(b)} First derivative of the heat capacity with respect to the rewiring probability $p_r$, shown for different generation levels $s$ (see legend).}
    \label{DGMRew}
\end{figure}

We now move to the case of DGM networks under rewiring, as shown in Fig.~\ref{DGMRew}. Finite-size scaling analyses confirm the existence of a finite critical rewiring probability in the thermodynamic limit, estimated at $p_{r,c}=0.05(3)$. Specifically, we observe that assuming an asymptotic limit of $p_{r,c}^\infty=0.03$ yields a consistent convergence to zero as system size increases. In contrast, setting $p_{r,c}^\infty=0.02$ results in an exponential decay with system size. This behavior establishes the attraction basin for this scale-invariant network against shortcut perturbations.

\begin{figure}[hbtp]
    \centering
    \includegraphics[width=0.75\linewidth]{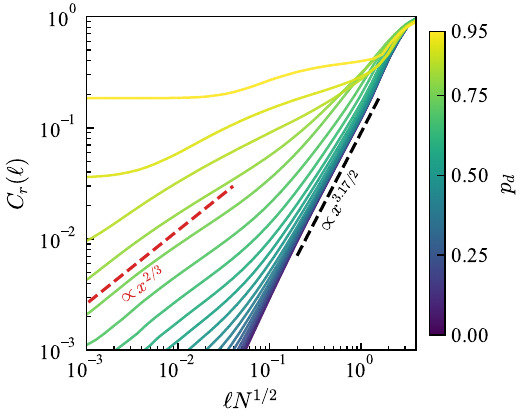}
    \caption{Scaling analysis of DGM networks. Correlation integral $C_{r}(\ell)$ as a function of the rescaled distance $\ell\sqrt{N}$ for networks at generation level $s = 10$, subjected to varying dilution probabilities (see legend). Dashed lines serve as guides to the eye, indicating slopes corresponding to correlation dimensions $D = 2/3$ (red dashed line) and $D = 3.17/2$ (blue dashed line).}
    \label{CorrDGM}
\end{figure}

Figure~\ref{CorrDGM} displays the correlation integral for different dilution probabilities in a DGM network of size $N = 29526$. In the low-dilution regime, the network exhibits a spectral dimension consistent with the expected theoretical value. However, as dilution increases, this dimension progressively deviates, eventually signaling a structural collapse.

\begin{figure}[H]
    \centering
    \includegraphics[width=1\linewidth]{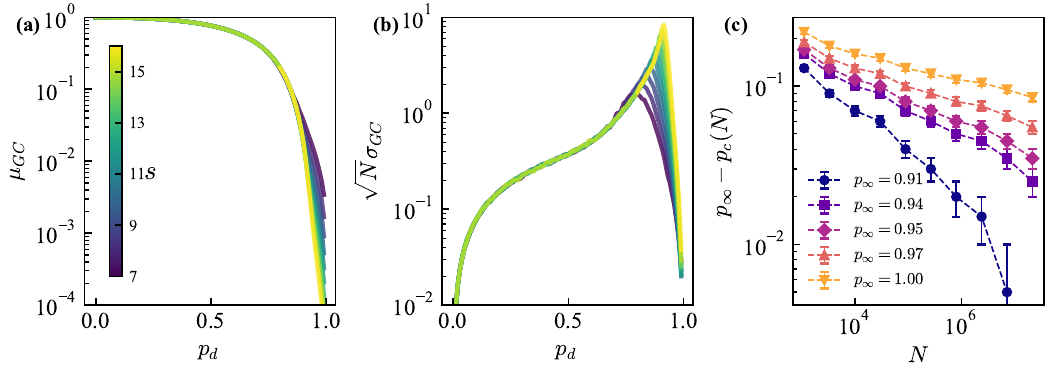}
    \caption{Percolation transition in DGM networks.
\textbf{(a)} Average fraction of nodes in the giant component as a function of dilution probability for varying generation levels $s$ (see legend).
\textbf{(b)} Standard deviation of the giant component fraction, representing the order parameter fluctuations across different levels $s$.
\textbf{(c)} Finite-size scaling analysis of the percolation threshold $p_c(s)$ versus system size.}
    \label{DGM}
\end{figure}

Figure~\ref{DGM} reports the percolation properties of diluted DGM networks. The left panel displays the average fraction of nodes belonging to the giant component as a function of dilution probability for varying hierarchical levels $s$ (see legend). As dilution increases, the giant component progressively decreases, suggesting a percolation threshold approaching one. Complementarily, the standard deviation of the percolation order parameter highlights the fluctuations associated with the phase transition. The peak of these fluctuations identifies the true critical point at $p_c=0.95(5)$. Furthermore, finite-size scaling analysis across different values of $s$ confirms that the critical threshold remains strictly below one.

\section{Tree networks}
\label{RTApp}
\begin{figure}[H]
    \centering
    \includegraphics[width=1\linewidth]{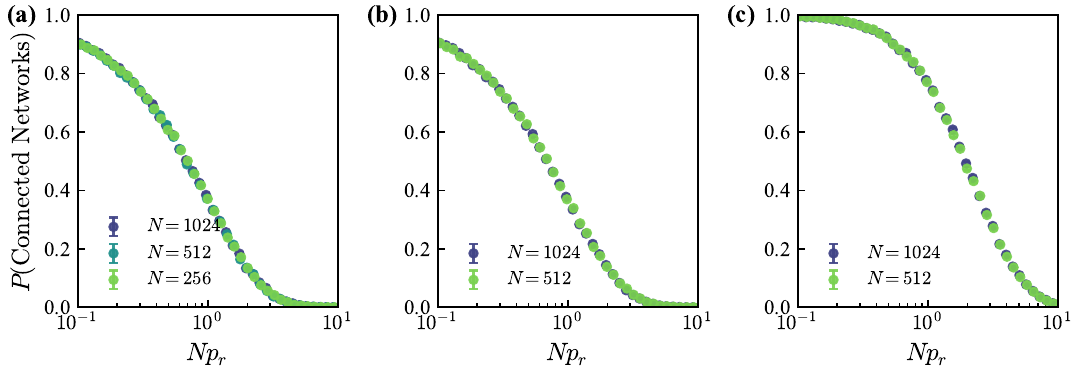}
    \caption{Connectivity transition in loop-free topologies. Probability of observing a fully connected network as a function of the rescaled rewiring probability $p_r N$.
\textbf{(a)} Barab\'asi--Albert networks with $m = 1$.
\textbf{(b)} Random trees.
\textbf{(c)} Ring graphs (1D lattices).
Different curves correspond to different system sizes $N$ (see legend).}
    \label{NP_r}
\end{figure}

Here, we report the percolation transition as a function of rewiring probability for random trees and Barab\'asi–Albert networks. Structurally, these systems are analogous to 1D rings because they lack loops. Notably, in all three cases, the critical probability required to sustain a giant connected component vanishes in the thermodynamic limit. This behavior is confirmed by the data collapse observed when the curves are properly rescaled with $N$, as illustrated in Fig.~\ref{NP_r}.

\section{Hierarchic-Modular networks}
\label{HMNApp}
\begin{figure}[H]
    \centering
    \includegraphics[width=1\linewidth]{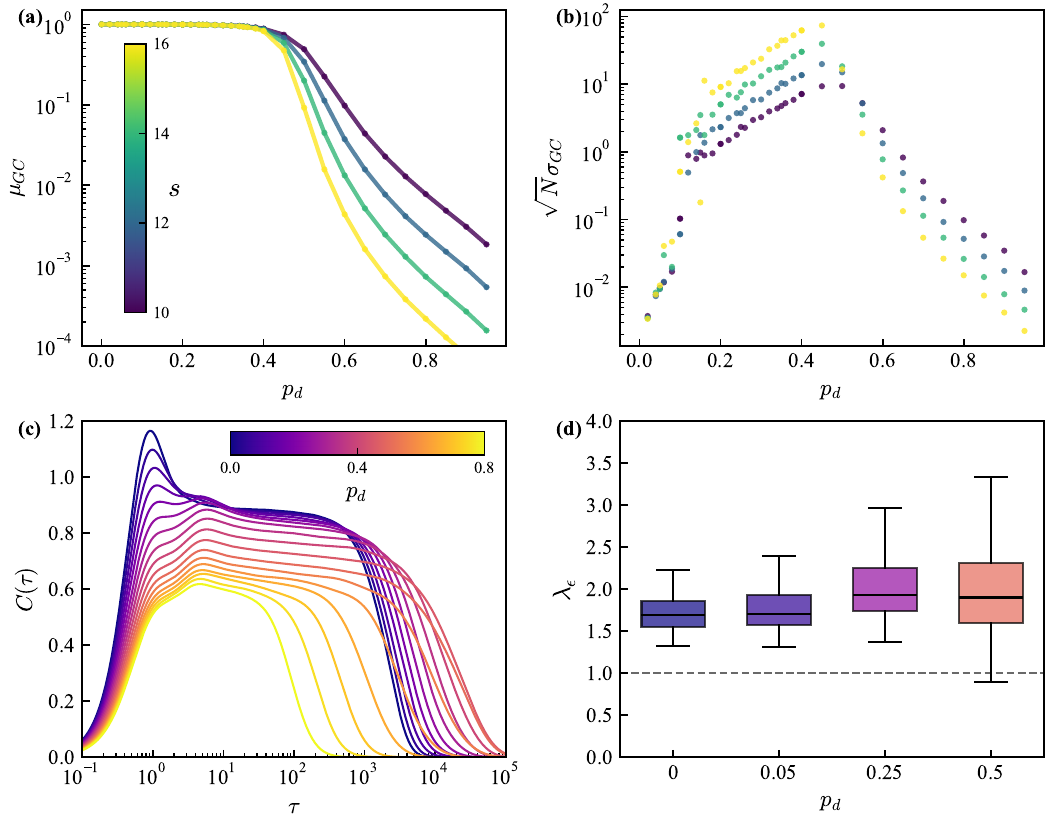}
    \caption{Percolation and heat capacity analysis in diluted HMNs.
\textbf{(a)} Average fraction of nodes in the giant component as a function of dilution probability $p_d$. Different curves correspond to different hierarchical levels $s$ (see legend).
\textbf{(b)} Standard deviation of the giant component fraction, representing order parameter fluctuations versus $p_d$ for different levels $s$.
\textbf{(c)} Heat capacity $C$ versus diffusion time $\tau$ for varying dilution probabilities (see colorbar).
\textbf{(d)} Lacunarity index, $\lambda_{\epsilon}$, calculated at a fixed scale $\epsilon=10^{-2}$, for diluted HMN with $m_{0}=3$, $\alpha=4$ and $s=11$. The horizontal dashed line marks the theoretical reference $\lambda_{\epsilon}=1$.}
    \label{HMN}
\end{figure}

Finally, Figure~\ref{HMN} illustrates the percolation behavior of diluted Hierarchical-Modular Networks (HMNs) across varying hierarchical levels $s$. The left panel depicts the average fraction of nodes in the giant component as a function of dilution probability. As dilution increases, the network undergoes a complex percolation transition. Crucially, the analysis of the susceptibility (defined as the standard deviation of the order parameter, see center panel) reveals a broad critical region, with fluctuations that appear to diverge in the thermodynamic limit even at low values of $p_d$. This suggests the presence of persistent critical fluctuations associated with the network's specific lacunarity across the entire dilution range. Finally, the right panel reports the heat capacity $C$ as a function of diffusion time $\tau$ for varying dilution probabilities (see colorbar).\\

\def\url#1{}

\begin{thebibliography}{47}%
\makeatletter
\providecommand \@ifxundefined [1]{%
 \@ifx{#1\undefined}
}%
\providecommand \@ifnum [1]{%
 \ifnum #1\expandafter \@firstoftwo
 \else \expandafter \@secondoftwo
 \fi
}%
\providecommand \@ifx [1]{%
 \ifx #1\expandafter \@firstoftwo
 \else \expandafter \@secondoftwo
 \fi
}%
\providecommand \natexlab [1]{#1}%
\providecommand \enquote  [1]{``#1''}%
\providecommand \bibnamefont  [1]{#1}%
\providecommand \bibfnamefont [1]{#1}%
\providecommand \citenamefont [1]{#1}%
\providecommand \href@noop [0]{\@secondoftwo}%
\providecommand \href [0]{\begingroup \@sanitize@url \@href}%
\providecommand \@href[1]{\@@startlink{#1}\@@href}%
\providecommand \@@href[1]{\endgroup#1\@@endlink}%
\providecommand \@sanitize@url [0]{\catcode `\\12\catcode `\$12\catcode
  `\&12\catcode `\#12\catcode `\^12\catcode `\_12\catcode `\%12\relax}%
\providecommand \@@startlink[1]{}%
\providecommand \@@endlink[0]{}%
\providecommand \url  [0]{\begingroup\@sanitize@url \@url }%
\providecommand \@url [1]{\endgroup\@href {#1}{\urlprefix }}%
\providecommand \urlprefix  [0]{URL }%
\providecommand \Eprint [0]{\href }%
\providecommand \doibase [0]{https://doi.org/}%
\providecommand \selectlanguage [0]{\@gobble}%
\providecommand \bibinfo  [0]{\@secondoftwo}%
\providecommand \bibfield  [0]{\@secondoftwo}%
\providecommand \translation [1]{[#1]}%
\providecommand \BibitemOpen [0]{}%
\providecommand \bibitemStop [0]{}%
\providecommand \bibitemNoStop [0]{.\EOS\space}%
\providecommand \EOS [0]{\spacefactor3000\relax}%
\providecommand \BibitemShut  [1]{\csname bibitem#1\endcsname}%
\let\auto@bib@innerbib\@empty
\bibitem [{\citenamefont {Betzel}\ and\ \citenamefont
  {Bassett}(2018)}]{Betzel2018}%
  \BibitemOpen
  \bibfield  {author} {\bibinfo {author} {\bibfnamefont {R.~F.}\ \bibnamefont
  {Betzel}}\ and\ \bibinfo {author} {\bibfnamefont {D.~S.}\ \bibnamefont
  {Bassett}},\ }\bibfield  {title} {\bibinfo {title} {Specificity and
  robustness of long-distance connections in weighted, interareal
  connectomes},\ }\href {https://doi.org/10.1073/pnas.1720186115} {\bibfield
  {journal} {\bibinfo  {journal} {Proc. Natl. Acad. Sci. U. S. A.}\ }\textbf
  {\bibinfo {volume} {115}},\ \bibinfo {pages} {E4880} (\bibinfo {year}
  {2018})}\BibitemShut {NoStop}%
\bibitem [{\citenamefont {Martin}\ \emph {et~al.}(2015)\citenamefont {Martin},
  \citenamefont {McGovern}, \citenamefont {Orozco} \emph
  {et~al.}}]{Martin2015}%
  \BibitemOpen
  \bibfield  {author} {\bibinfo {author} {\bibfnamefont {P.}~\bibnamefont
  {Martin}}, \bibinfo {author} {\bibfnamefont {A.}~\bibnamefont {McGovern}},
  \bibinfo {author} {\bibfnamefont {G.}~\bibnamefont {Orozco}}, \emph
  {et~al.},\ }\bibfield  {title} {\bibinfo {title} {Capture hi-c reveals novel
  candidate genes and complex long-range interactions with related autoimmune
  risk loci},\ }\href {https://doi.org/10.1038/ncomms10069} {\bibfield
  {journal} {\bibinfo  {journal} {Nat. Commun.}\ }\textbf {\bibinfo {volume}
  {6}},\ \bibinfo {pages} {10069} (\bibinfo {year} {2015})}\BibitemShut
  {NoStop}%
\bibitem [{\citenamefont {Simonov}\ and\ \citenamefont
  {Goodwin}(2020)}]{Simonov2020}%
  \BibitemOpen
  \bibfield  {author} {\bibinfo {author} {\bibfnamefont {A.}~\bibnamefont
  {Simonov}}\ and\ \bibinfo {author} {\bibfnamefont {A.~L.}\ \bibnamefont
  {Goodwin}},\ }\bibfield  {title} {\bibinfo {title} {Designing disorder into
  crystalline materials},\ }\href {https://doi.org/10.1038/s41570-020-00228-3}
  {\bibfield  {journal} {\bibinfo  {journal} {Nat. Rev. Chem.}\ }\textbf
  {\bibinfo {volume} {4}},\ \bibinfo {pages} {657} (\bibinfo {year}
  {2020})}\BibitemShut {NoStop}%
\bibitem [{\citenamefont {Sporns}(2010)}]{SpornsBook}%
  \BibitemOpen
  \bibfield  {author} {\bibinfo {author} {\bibfnamefont {O.}~\bibnamefont
  {Sporns}},\ }\href {https://doi.org/10.7551/mitpress/8476.001.0001} {\emph
  {\bibinfo {title} {Networks of the Brain}}}\ (\bibinfo  {publisher} {The MIT
  Press},\ \bibinfo {year} {2010})\BibitemShut {NoStop}%
\bibitem [{\citenamefont {Kaiser}\ and\ \citenamefont
  {Hilgetag}(2006)}]{Kaiser2006}%
  \BibitemOpen
  \bibfield  {author} {\bibinfo {author} {\bibfnamefont {M.}~\bibnamefont
  {Kaiser}}\ and\ \bibinfo {author} {\bibfnamefont {C.~C.}\ \bibnamefont
  {Hilgetag}},\ }\bibfield  {title} {\bibinfo {title} {Nonoptimal component
  placement, but short processing paths, due to long-distance projections in
  neural systems},\ }\href {https://doi.org/10.1371/journal.pcbi.0020095}
  {\bibfield  {journal} {\bibinfo  {journal} {PLoS Comp. Biol.}\ }\textbf
  {\bibinfo {volume} {2}},\ \bibinfo {pages} {e95} (\bibinfo {year}
  {2006})}\BibitemShut {NoStop}%
\bibitem [{\citenamefont {Bullmore}\ and\ \citenamefont
  {Sporns}(2009)}]{Bullmore2009}%
  \BibitemOpen
  \bibfield  {author} {\bibinfo {author} {\bibfnamefont {E.}~\bibnamefont
  {Bullmore}}\ and\ \bibinfo {author} {\bibfnamefont {O.}~\bibnamefont
  {Sporns}},\ }\bibfield  {title} {\bibinfo {title} {Complex brain networks:
  graph theoretical analysis of structural and functional systems},\ }\href
  {https://doi.org/10.1038/nrn2575} {\bibfield  {journal} {\bibinfo  {journal}
  {Nat. Rev. Neurosci.}\ }\textbf {\bibinfo {volume} {10}},\ \bibinfo {pages}
  {186} (\bibinfo {year} {2009})}\BibitemShut {NoStop}%
\bibitem [{\citenamefont {Betzel}\ and\ \citenamefont
  {Bassett}(2017)}]{Betzel2017}%
  \BibitemOpen
  \bibfield  {author} {\bibinfo {author} {\bibfnamefont {R.~F.}\ \bibnamefont
  {Betzel}}\ and\ \bibinfo {author} {\bibfnamefont {D.~S.}\ \bibnamefont
  {Bassett}},\ }\bibfield  {title} {\bibinfo {title} {Multi-scale brain
  networks},\ }\href {https://doi.org/10.1016/j.neuroimage.2016.11.006}
  {\bibfield  {journal} {\bibinfo  {journal} {NeuroImage}\ }\textbf {\bibinfo
  {volume} {160}},\ \bibinfo {pages} {73} (\bibinfo {year} {2017})},\ \bibinfo
  {note} {functional Architecture of the Brain}\BibitemShut {NoStop}%
\bibitem [{\citenamefont {Schein}\ and\ \citenamefont
  {Aichelburg}(1996)}]{Wormholes}%
  \BibitemOpen
  \bibfield  {author} {\bibinfo {author} {\bibfnamefont {F.}~\bibnamefont
  {Schein}}\ and\ \bibinfo {author} {\bibfnamefont {P.~C.}\ \bibnamefont
  {Aichelburg}},\ }\bibfield  {title} {\bibinfo {title} {Traversable wormholes
  in geometries of charged shells},\ }\href
  {https://doi.org/10.1103/PhysRevLett.77.4130} {\bibfield  {journal} {\bibinfo
   {journal} {Phys. Rev. Lett.}\ }\textbf {\bibinfo {volume} {77}},\ \bibinfo
  {pages} {4130} (\bibinfo {year} {1996})}\BibitemShut {NoStop}%
\bibitem [{\citenamefont {Chen}\ \emph {et~al.}(2014)\citenamefont {Chen},
  \citenamefont {Huang},\ and\ \citenamefont {Lai}}]{Chen2014}%
  \BibitemOpen
  \bibfield  {author} {\bibinfo {author} {\bibfnamefont {Y.-C.}\ \bibnamefont
  {Chen}}, \bibinfo {author} {\bibfnamefont {Z.-G.}\ \bibnamefont {Huang}},\
  and\ \bibinfo {author} {\bibfnamefont {Y.-C.}\ \bibnamefont {Lai}},\
  }\bibfield  {title} {\bibinfo {title} {Controlling extreme events on complex
  networks},\ }\href {https://doi.org/10.1038/srep06121} {\bibfield  {journal}
  {\bibinfo  {journal} {Scientific Reports}\ }\textbf {\bibinfo {volume} {4}},\
  \bibinfo {pages} {6121} (\bibinfo {year} {2014})}\BibitemShut {NoStop}%
\bibitem [{\citenamefont {Morris}\ and\ \citenamefont {{\v
  C}ejka}(2015)}]{Morris2015}%
  \BibitemOpen
  \bibfield  {author} {\bibinfo {author} {\bibfnamefont {R.~E.}\ \bibnamefont
  {Morris}}\ and\ \bibinfo {author} {\bibfnamefont {J.}~\bibnamefont {{\v
  C}ejka}},\ }\bibfield  {title} {\bibinfo {title} {Exploiting chemically
  selective weakness in solids as a route to new porous materials},\ }\href
  {https://doi.org/10.1038/nchem.2220} {\bibfield  {journal} {\bibinfo
  {journal} {Nat. Chem.}\ }\textbf {\bibinfo {volume} {7}},\ \bibinfo {pages}
  {381} (\bibinfo {year} {2015})}\BibitemShut {NoStop}%
\bibitem [{\citenamefont {Roth}\ and\ \citenamefont {et~al.}(2013)}]{Roth2013}%
  \BibitemOpen
  \bibfield  {author} {\bibinfo {author} {\bibfnamefont {W.~J.}\ \bibnamefont
  {Roth}}\ and\ \bibinfo {author} {\bibnamefont {et~al.}},\ }\bibfield  {title}
  {\bibinfo {title} {A family of zeolites with controlled pore size prepared
  using a top-down method},\ }\href {https://doi.org/10.1038/nchem.1675}
  {\bibfield  {journal} {\bibinfo  {journal} {Nat. Chem.}\ }\textbf {\bibinfo
  {volume} {5}},\ \bibinfo {pages} {628} (\bibinfo {year} {2013})}\BibitemShut
  {NoStop}%
\bibitem [{\citenamefont {Chen}(2016)}]{Chen2016}%
  \BibitemOpen
  \bibfield  {author} {\bibinfo {author} {\bibfnamefont {X.~e.~a.}\
  \bibnamefont {Chen}},\ }\bibfield  {title} {\bibinfo {title} {Oxygen
  vacancies and their impact on electronic phases in vanadium oxides},\ }\href
  {https://doi.org/10.1103/PhysRevApplied.7.034008} {\bibfield  {journal}
  {\bibinfo  {journal} {Phys. Rev. Applied}\ }\textbf {\bibinfo {volume} {7}},\
  \bibinfo {pages} {034008} (\bibinfo {year} {2016})}\BibitemShut {NoStop}%
\bibitem [{\citenamefont {Sun}(2021)}]{Sun2021}%
  \BibitemOpen
  \bibfield  {author} {\bibinfo {author} {\bibfnamefont {Y.~e.~a.}\
  \bibnamefont {Sun}},\ }\bibfield  {title} {\bibinfo {title} {Anion vacancy
  effects on structural stability in oxynitrides},\ }\href
  {https://doi.org/10.1039/D1MA00122A} {\bibfield  {journal} {\bibinfo
  {journal} {Mater. Adv.}\ }\textbf {\bibinfo {volume} {2}},\ \bibinfo {pages}
  {1234} (\bibinfo {year} {2021})}\BibitemShut {NoStop}%
\bibitem [{\citenamefont {Zhou}(2023)}]{Zhou2023}%
  \BibitemOpen
  \bibfield  {author} {\bibinfo {author} {\bibfnamefont {J.~e.~a.}\
  \bibnamefont {Zhou}},\ }\bibfield  {title} {\bibinfo {title} {Vacancy defects
  in skyrmion crystals: structural and magnetic effects},\ }\href
  {https://doi.org/10.1016/j.jmmm.2023.170147} {\bibfield  {journal} {\bibinfo
  {journal} {J. Magn. Magn. Mater.}\ }\textbf {\bibinfo {volume} {562}},\
  \bibinfo {pages} {170147} (\bibinfo {year} {2023})}\BibitemShut {NoStop}%
\bibitem [{\citenamefont {Watts}\ and\ \citenamefont
  {Strogatz}(1998)}]{Watts1998}%
  \BibitemOpen
  \bibfield  {author} {\bibinfo {author} {\bibfnamefont {D.~J.}\ \bibnamefont
  {Watts}}\ and\ \bibinfo {author} {\bibfnamefont {S.~H.}\ \bibnamefont
  {Strogatz}},\ }\bibfield  {title} {\bibinfo {title} {Collective dynamics of
  ‘small-world’ networks},\ }\href {https://doi.org/10.1038/30918}
  {\bibfield  {journal} {\bibinfo  {journal} {Nature}\ }\textbf {\bibinfo
  {volume} {393}},\ \bibinfo {pages} {440} (\bibinfo {year}
  {1998})}\BibitemShut {NoStop}%
\bibitem [{\citenamefont {Newman}\ and\ \citenamefont
  {Watts}(1999)}]{Newman1999}%
  \BibitemOpen
  \bibfield  {author} {\bibinfo {author} {\bibfnamefont {M.}~\bibnamefont
  {Newman}}\ and\ \bibinfo {author} {\bibfnamefont {D.}~\bibnamefont {Watts}},\
  }\bibfield  {title} {\bibinfo {title} {Renormalization group analysis of the
  small-world network model},\ }\href
  {https://doi.org/https://doi.org/10.1016/S0375-9601(99)00757-4} {\bibfield
  {journal} {\bibinfo  {journal} {Phys. Lett. A}\ }\textbf {\bibinfo {volume}
  {263}},\ \bibinfo {pages} {341} (\bibinfo {year} {1999})}\BibitemShut
  {NoStop}%
\bibitem [{\citenamefont {Newman}(2000)}]{Newman2000}%
  \BibitemOpen
  \bibfield  {author} {\bibinfo {author} {\bibfnamefont {M.~E.~J.}\
  \bibnamefont {Newman}},\ }\bibfield  {title} {\bibinfo {title} {Models of the
  small world},\ }\href {https://doi.org/10.1023/A:1026485807148} {\bibfield
  {journal} {\bibinfo  {journal} {J. Stat. Phys.}\ }\textbf {\bibinfo {volume}
  {101}},\ \bibinfo {pages} {819} (\bibinfo {year} {2000})}\BibitemShut
  {NoStop}%
\bibitem [{\citenamefont {Mill\'an}\ \emph {et~al.}(2021)\citenamefont
  {Mill\'an}, \citenamefont {Gori}, \citenamefont {Battiston}, \citenamefont
  {Enss},\ and\ \citenamefont {Defenu}}]{Millan2021}%
  \BibitemOpen
  \bibfield  {author} {\bibinfo {author} {\bibfnamefont {A.~P.}\ \bibnamefont
  {Mill\'an}}, \bibinfo {author} {\bibfnamefont {G.}~\bibnamefont {Gori}},
  \bibinfo {author} {\bibfnamefont {F.}~\bibnamefont {Battiston}}, \bibinfo
  {author} {\bibfnamefont {T.}~\bibnamefont {Enss}},\ and\ \bibinfo {author}
  {\bibfnamefont {N.}~\bibnamefont {Defenu}},\ }\bibfield  {title} {\bibinfo
  {title} {Complex networks with tuneable spectral dimension as a universality
  playground},\ }\href {https://doi.org/10.1103/PhysRevResearch.3.023015}
  {\bibfield  {journal} {\bibinfo  {journal} {Phys. Rev. Res.}\ }\textbf
  {\bibinfo {volume} {3}},\ \bibinfo {pages} {023015} (\bibinfo {year}
  {2021})}\BibitemShut {NoStop}%
\bibitem [{\citenamefont {Vojta}(2006)}]{Vojta2006}%
  \BibitemOpen
  \bibfield  {author} {\bibinfo {author} {\bibfnamefont {T.}~\bibnamefont
  {Vojta}},\ }\bibfield  {title} {\bibinfo {title} {Rare region effects at
  classical, quantum and nonequilibrium phase transitions},\ }\href
  {https://doi.org/10.1088/0305-4470/39/22/R01} {\bibfield  {journal} {\bibinfo
   {journal} {J. Phys. A Math. Gen.}\ }\textbf {\bibinfo {volume} {39}},\
  \bibinfo {pages} {R143} (\bibinfo {year} {2006})}\BibitemShut {NoStop}%
\bibitem [{\citenamefont {Griffiths}(1969)}]{Griffiths1969}%
  \BibitemOpen
  \bibfield  {author} {\bibinfo {author} {\bibfnamefont {R.~B.}\ \bibnamefont
  {Griffiths}},\ }\bibfield  {title} {\bibinfo {title} {Nonanalytic behavior
  above the critical point in a random ising ferromagnet},\ }\href
  {https://doi.org/10.1103/PhysRevLett.23.17} {\bibfield  {journal} {\bibinfo
  {journal} {Phys. Rev. Lett.}\ }\textbf {\bibinfo {volume} {23}},\ \bibinfo
  {pages} {17} (\bibinfo {year} {1969})}\BibitemShut {NoStop}%
\bibitem [{\citenamefont {Westphal}\ \emph {et~al.}(1992)\citenamefont
  {Westphal}, \citenamefont {Kleemann},\ and\ \citenamefont
  {Glinchuk}}]{Westphal1992}%
  \BibitemOpen
  \bibfield  {author} {\bibinfo {author} {\bibfnamefont {V.}~\bibnamefont
  {Westphal}}, \bibinfo {author} {\bibfnamefont {W.}~\bibnamefont {Kleemann}},\
  and\ \bibinfo {author} {\bibfnamefont {M.~D.}\ \bibnamefont {Glinchuk}},\
  }\bibfield  {title} {\bibinfo {title} {Diffuse phase transitions and
  random-field-induced domain states of the ``relaxor'' ferroelectric
  ${\mathrm{pbmg}}_{1/3}$${\mathrm{nb}}_{2/3}$${\mathrm{o}}_{3}$},\ }\href
  {https://doi.org/10.1103/PhysRevLett.68.847} {\bibfield  {journal} {\bibinfo
  {journal} {Phys. Rev. Lett.}\ }\textbf {\bibinfo {volume} {68}},\ \bibinfo
  {pages} {847} (\bibinfo {year} {1992})}\BibitemShut {NoStop}%
\bibitem [{\citenamefont {Timonin}(1997)}]{Timonin1997}%
  \BibitemOpen
  \bibfield  {author} {\bibinfo {author} {\bibfnamefont {P.}~\bibnamefont
  {Timonin}},\ }\bibfield  {title} {\bibinfo {title} {Griffiths' phase in
  dilute ferroelectrics},\ }\href@noop {} {\bibfield  {journal} {\bibinfo
  {journal} {Ferroelectrics}\ }\textbf {\bibinfo {volume} {199}},\ \bibinfo
  {pages} {69} (\bibinfo {year} {1997})}\BibitemShut {NoStop}%
\bibitem [{\citenamefont {Villegas}\ \emph {et~al.}(2023)\citenamefont
  {Villegas}, \citenamefont {Gili}, \citenamefont {Caldarelli},\ and\
  \citenamefont {Gabrielli}}]{LRG}%
  \BibitemOpen
  \bibfield  {author} {\bibinfo {author} {\bibfnamefont {P.}~\bibnamefont
  {Villegas}}, \bibinfo {author} {\bibfnamefont {T.}~\bibnamefont {Gili}},
  \bibinfo {author} {\bibfnamefont {G.}~\bibnamefont {Caldarelli}},\ and\
  \bibinfo {author} {\bibfnamefont {A.}~\bibnamefont {Gabrielli}},\ }\bibfield
  {title} {\bibinfo {title} {Laplacian renormalization group for heterogeneous
  networks},\ }\href {https://doi.org/10.1038/s41567-022-01866-8} {\bibfield
  {journal} {\bibinfo  {journal} {Nat. Phys.}\ }\textbf {\bibinfo {volume}
  {19}},\ \bibinfo {pages} {445} (\bibinfo {year} {2023})}\BibitemShut
  {NoStop}%
\bibitem [{\citenamefont {Poggialini}\ \emph {et~al.}(2025)\citenamefont
  {Poggialini}, \citenamefont {Villegas}, \citenamefont {Mu\~noz},\ and\
  \citenamefont {Gabrielli}}]{Poggialini2025}%
  \BibitemOpen
  \bibfield  {author} {\bibinfo {author} {\bibfnamefont {A.}~\bibnamefont
  {Poggialini}}, \bibinfo {author} {\bibfnamefont {P.}~\bibnamefont
  {Villegas}}, \bibinfo {author} {\bibfnamefont {M.~A.}\ \bibnamefont
  {Mu\~noz}},\ and\ \bibinfo {author} {\bibfnamefont {A.}~\bibnamefont
  {Gabrielli}},\ }\bibfield  {title} {\bibinfo {title} {Networks with many
  structural scales: A renormalization group perspective},\ }\href
  {https://doi.org/10.1103/PhysRevLett.134.057401} {\bibfield  {journal}
  {\bibinfo  {journal} {Phys. Rev. Lett.}\ }\textbf {\bibinfo {volume} {134}},\
  \bibinfo {pages} {057401} (\bibinfo {year} {2025})}\BibitemShut {NoStop}%
\bibitem [{\citenamefont {Cassi}(1996)}]{Cassi1996}%
  \BibitemOpen
  \bibfield  {author} {\bibinfo {author} {\bibfnamefont {D.}~\bibnamefont
  {Cassi}},\ }\bibfield  {title} {\bibinfo {title} {Local vs average behavior
  on inhomogeneous structures: Recurrence on the average and a further
  extension of mermin-wagner theorem on graphs},\ }\href
  {https://doi.org/10.1103/PhysRevLett.76.2941} {\bibfield  {journal} {\bibinfo
   {journal} {Phys. Rev. Lett.}\ }\textbf {\bibinfo {volume} {76}},\ \bibinfo
  {pages} {2941} (\bibinfo {year} {1996})}\BibitemShut {NoStop}%
\bibitem [{\citenamefont {Cassi}(1992)}]{Cassi1992}%
  \BibitemOpen
  \bibfield  {author} {\bibinfo {author} {\bibfnamefont {D.}~\bibnamefont
  {Cassi}},\ }\bibfield  {title} {\bibinfo {title} {Phase transitions and
  random walks on graphs: A generalization of the mermin-wagner theorem to
  disordered lattices, fractals, and other discrete structures},\ }\href
  {https://doi.org/10.1103/PhysRevLett.68.3631} {\bibfield  {journal} {\bibinfo
   {journal} {Phys. Rev. Lett.}\ }\textbf {\bibinfo {volume} {68}},\ \bibinfo
  {pages} {3631} (\bibinfo {year} {1992})}\BibitemShut {NoStop}%
\bibitem [{\citenamefont {Villegas}(2025)}]{Villegas2025}%
  \BibitemOpen
  \bibfield  {author} {\bibinfo {author} {\bibfnamefont {P.}~\bibnamefont
  {Villegas}},\ }\bibfield  {title} {\bibinfo {title} {Strange attractors in
  complex networks},\ }\href {https://doi.org/10.1103/PhysRevE.111.L042301}
  {\bibfield  {journal} {\bibinfo  {journal} {Phys. Rev. E}\ }\textbf {\bibinfo
  {volume} {111}},\ \bibinfo {pages} {L042301} (\bibinfo {year}
  {2025})}\BibitemShut {NoStop}%
\bibitem [{\citenamefont {Gabrielli}\ \emph {et~al.}(2025)\citenamefont
  {Gabrielli}, \citenamefont {Garlaschelli}, \citenamefont {Patil} \emph
  {et~al.}}]{Gabrielli2025}%
  \BibitemOpen
  \bibfield  {author} {\bibinfo {author} {\bibfnamefont {A.}~\bibnamefont
  {Gabrielli}}, \bibinfo {author} {\bibfnamefont {D.}~\bibnamefont
  {Garlaschelli}}, \bibinfo {author} {\bibfnamefont {S.}~\bibnamefont {Patil}},
  \emph {et~al.},\ }\bibfield  {title} {\bibinfo {title} {Network
  renormalization},\ }\href {https://doi.org/10.1038/s42254-025-00817-5}
  {\bibfield  {journal} {\bibinfo  {journal} {Nat. Rev. Phys.}\ }\textbf
  {\bibinfo {volume} {7}},\ \bibinfo {pages} {203} (\bibinfo {year}
  {2025})}\BibitemShut {NoStop}%
\bibitem [{\citenamefont {De~Domenico}\ and\ \citenamefont
  {Biamonte}(2016)}]{Domenico2016}%
  \BibitemOpen
  \bibfield  {author} {\bibinfo {author} {\bibfnamefont {M.}~\bibnamefont
  {De~Domenico}}\ and\ \bibinfo {author} {\bibfnamefont {J.}~\bibnamefont
  {Biamonte}},\ }\bibfield  {title} {\bibinfo {title} {Spectral entropies as
  information-theoretic tools for complex network comparison},\ }\href
  {https://doi.org/10.1103/PhysRevX.6.041062} {\bibfield  {journal} {\bibinfo
  {journal} {Phys. Rev. X}\ }\textbf {\bibinfo {volume} {6}},\ \bibinfo {pages}
  {041062} (\bibinfo {year} {2016})}\BibitemShut {NoStop}%
\bibitem [{\citenamefont {Villegas}\ \emph {et~al.}(2022)\citenamefont
  {Villegas}, \citenamefont {Gabrielli}, \citenamefont {Santucci},
  \citenamefont {Caldarelli},\ and\ \citenamefont {Gili}}]{InfoCore}%
  \BibitemOpen
  \bibfield  {author} {\bibinfo {author} {\bibfnamefont {P.}~\bibnamefont
  {Villegas}}, \bibinfo {author} {\bibfnamefont {A.}~\bibnamefont {Gabrielli}},
  \bibinfo {author} {\bibfnamefont {F.}~\bibnamefont {Santucci}}, \bibinfo
  {author} {\bibfnamefont {G.}~\bibnamefont {Caldarelli}},\ and\ \bibinfo
  {author} {\bibfnamefont {T.}~\bibnamefont {Gili}},\ }\bibfield  {title}
  {\bibinfo {title} {Laplacian paths in complex networks: Information core
  emerges from entropic transitions},\ }\href
  {https://doi.org/10.1103/PhysRevResearch.4.033196} {\bibfield  {journal}
  {\bibinfo  {journal} {Phys. Rev. Res.}\ }\textbf {\bibinfo {volume} {4}},\
  \bibinfo {pages} {033196} (\bibinfo {year} {2022})}\BibitemShut {NoStop}%
\bibitem [{\citenamefont {Falsi}\ \emph {et~al.}(2025)\citenamefont {Falsi},
  \citenamefont {Villegas}, \citenamefont {Gili}, \citenamefont {Agranat},\
  and\ \citenamefont {DelRe}}]{TBM}%
  \BibitemOpen
  \bibfield  {author} {\bibinfo {author} {\bibfnamefont {L.}~\bibnamefont
  {Falsi}}, \bibinfo {author} {\bibfnamefont {P.}~\bibnamefont {Villegas}},
  \bibinfo {author} {\bibfnamefont {T.}~\bibnamefont {Gili}}, \bibinfo {author}
  {\bibfnamefont {A.~J.}\ \bibnamefont {Agranat}},\ and\ \bibinfo {author}
  {\bibfnamefont {E.}~\bibnamefont {DelRe}},\ }\bibfield  {title} {\bibinfo
  {title} {Topological protection breakdown: A route to frustrated
  ferroelectricity},\ }\href {https://doi.org/10.1103/h6j7-cgwz} {\bibfield
  {journal} {\bibinfo  {journal} {Phys. Rev. Res.}\ }\textbf {\bibinfo {volume}
  {7}},\ \bibinfo {pages} {043038} (\bibinfo {year} {2025})}\BibitemShut
  {NoStop}%
\bibitem [{\citenamefont {Donetti}\ and\ \citenamefont
  {Destri}(2004)}]{Donetti2004}%
  \BibitemOpen
  \bibfield  {author} {\bibinfo {author} {\bibfnamefont {L.}~\bibnamefont
  {Donetti}}\ and\ \bibinfo {author} {\bibfnamefont {C.}~\bibnamefont
  {Destri}},\ }\bibfield  {title} {\bibinfo {title} {The statistical geometry
  of scale-free random trees},\ }\href
  {https://doi.org/10.1088/0305-4470/37/23/004} {\bibfield  {journal} {\bibinfo
   {journal} {J. Phys. A Math. Gen.}\ }\textbf {\bibinfo {volume} {37}},\
  \bibinfo {pages} {6003} (\bibinfo {year} {2004})}\BibitemShut {NoStop}%
\bibitem [{\citenamefont {Rozenfeld}\ \emph {et~al.}(2007)\citenamefont
  {Rozenfeld}, \citenamefont {Havlin},\ and\ \citenamefont
  {Ben-Avraham}}]{Rozenfeld2007}%
  \BibitemOpen
  \bibfield  {author} {\bibinfo {author} {\bibfnamefont {H.~D.}\ \bibnamefont
  {Rozenfeld}}, \bibinfo {author} {\bibfnamefont {S.}~\bibnamefont {Havlin}},\
  and\ \bibinfo {author} {\bibfnamefont {D.}~\bibnamefont {Ben-Avraham}},\
  }\bibfield  {title} {\bibinfo {title} {Fractal and transfractal recursive
  scale-free nets},\ }\href {https://doi.org/10.1088/1367-2630/9/6/175}
  {\bibfield  {journal} {\bibinfo  {journal} {New J. Phys.}\ }\textbf {\bibinfo
  {volume} {9}},\ \bibinfo {pages} {175} (\bibinfo {year} {2007})}\BibitemShut
  {NoStop}%
\bibitem [{\citenamefont {Dorogovtsev}\ \emph {et~al.}(2002)\citenamefont
  {Dorogovtsev}, \citenamefont {Goltsev},\ and\ \citenamefont
  {Mendes}}]{Dorogovtsev2002}%
  \BibitemOpen
  \bibfield  {author} {\bibinfo {author} {\bibfnamefont {S.~N.}\ \bibnamefont
  {Dorogovtsev}}, \bibinfo {author} {\bibfnamefont {A.~V.}\ \bibnamefont
  {Goltsev}},\ and\ \bibinfo {author} {\bibfnamefont {J.~F.~F.}\ \bibnamefont
  {Mendes}},\ }\bibfield  {title} {\bibinfo {title} {Pseudofractal scale-free
  web},\ }\href {https://doi.org/10.1103/PhysRevE.65.066122} {\bibfield
  {journal} {\bibinfo  {journal} {Phys. Rev. E}\ }\textbf {\bibinfo {volume}
  {65}},\ \bibinfo {pages} {066122} (\bibinfo {year} {2002})}\BibitemShut
  {NoStop}%
\bibitem [{\citenamefont {Holme}\ and\ \citenamefont {Kim}(2002)}]{KimHolme}%
  \BibitemOpen
  \bibfield  {author} {\bibinfo {author} {\bibfnamefont {P.}~\bibnamefont
  {Holme}}\ and\ \bibinfo {author} {\bibfnamefont {B.~J.}\ \bibnamefont
  {Kim}},\ }\bibfield  {title} {\bibinfo {title} {Growing scale-free networks
  with tunable clustering},\ }\href
  {https://doi.org/10.1103/PhysRevE.65.026107} {\bibfield  {journal} {\bibinfo
  {journal} {Phys. Rev. E}\ }\textbf {\bibinfo {volume} {65}},\ \bibinfo
  {pages} {026107} (\bibinfo {year} {2002})}\BibitemShut {NoStop}%
\bibitem [{\citenamefont {Moretti}\ and\ \citenamefont
  {Mu{\~n}oz}(2013)}]{Moretti2013}%
  \BibitemOpen
  \bibfield  {author} {\bibinfo {author} {\bibfnamefont {P.}~\bibnamefont
  {Moretti}}\ and\ \bibinfo {author} {\bibfnamefont {M.~A.}\ \bibnamefont
  {Mu{\~n}oz}},\ }\bibfield  {title} {\bibinfo {title} {Griffiths phases and
  the stretching of criticality in brain networks},\ }\href
  {https://doi.org/10.1038/ncomms3521} {\bibfield  {journal} {\bibinfo
  {journal} {Nat. Comm.}\ }\textbf {\bibinfo {volume} {4}},\ \bibinfo {pages}
  {2521} (\bibinfo {year} {2013})}\BibitemShut {NoStop}%
\bibitem [{SM()}]{SM}%
  \BibitemOpen
  \href@noop {} {}\bibinfo {note} {See Supplementary Videos at [link] supporting the findings in the main text.}\BibitemShut {Stop}%
\bibitem [{\citenamefont {{Alexander, S.}}\ and\ \citenamefont {{Orbach,
  R.}}(1982)}]{AO}%
  \BibitemOpen
  \bibfield  {author} {\bibinfo {author} {\bibnamefont {{Alexander, S.}}}\ and\
  \bibinfo {author} {\bibnamefont {{Orbach, R.}}},\ }\bibfield  {title}
  {\bibinfo {title} {Density of states on fractals : « fractons »},\ }\href
  {https://doi.org/10.1051/jphyslet:019820043017062500} {\bibfield  {journal}
  {\bibinfo  {journal} {J. Physique Lett.}\ }\textbf {\bibinfo {volume} {43}},\
  \bibinfo {pages} {625} (\bibinfo {year} {1982})}\BibitemShut {NoStop}%
\bibitem [{\citenamefont {Grassberger}\ and\ \citenamefont
  {Procaccia}(1983)}]{CorrDim}%
  \BibitemOpen
  \bibfield  {author} {\bibinfo {author} {\bibfnamefont {P.}~\bibnamefont
  {Grassberger}}\ and\ \bibinfo {author} {\bibfnamefont {I.}~\bibnamefont
  {Procaccia}},\ }\bibfield  {title} {\bibinfo {title} {Measuring the
  strangeness of strange attractors},\ }\href
  {https://doi.org/10.1016/0167-2789(83)90298-1} {\bibfield  {journal}
  {\bibinfo  {journal} {Phys. D: Nonlinear Phenom.}\ }\textbf {\bibinfo
  {volume} {9}},\ \bibinfo {pages} {189–208} (\bibinfo {year}
  {1983})}\BibitemShut {NoStop}%
\bibitem [{\citenamefont {Villegas}\ \emph {et~al.}(2024)\citenamefont
  {Villegas}, \citenamefont {Gili}, \citenamefont {Caldarelli},\ and\
  \citenamefont {Gabrielli}}]{ScaleFree}%
  \BibitemOpen
  \bibfield  {author} {\bibinfo {author} {\bibfnamefont {P.}~\bibnamefont
  {Villegas}}, \bibinfo {author} {\bibfnamefont {T.}~\bibnamefont {Gili}},
  \bibinfo {author} {\bibfnamefont {G.}~\bibnamefont {Caldarelli}},\ and\
  \bibinfo {author} {\bibfnamefont {A.}~\bibnamefont {Gabrielli}},\ }\bibfield
  {title} {\bibinfo {title} {Evidence of scale-free clusters of vegetation in
  tropical rainforests},\ }\href {https://doi.org/10.1103/PhysRevE.109.L042402}
  {\bibfield  {journal} {\bibinfo  {journal} {Phys. Rev. E}\ }\textbf {\bibinfo
  {volume} {109}},\ \bibinfo {pages} {L042402} (\bibinfo {year}
  {2024})}\BibitemShut {NoStop}%
\bibitem [{\citenamefont {Mandelbrot}(1983)}]{Mandelbrot}%
  \BibitemOpen
  \bibfield  {author} {\bibinfo {author} {\bibfnamefont {B.~B.}\ \bibnamefont
  {Mandelbrot}},\ }\href@noop {} {\emph {\bibinfo {title} {The Fractal Geometry
  of Nature}}},\ \bibinfo {edition} {revised and enlarged edition}\ ed.\
  (\bibinfo  {publisher} {W. H. Freeman and Co.},\ \bibinfo {address} {New
  York},\ \bibinfo {year} {1983})\ p.\ \bibinfo {pages} {495}\BibitemShut
  {NoStop}%
\bibitem [{\citenamefont {Gabrielli}\ \emph {et~al.}(2006)\citenamefont
  {Gabrielli}, \citenamefont {Labini}, \citenamefont {Joyce},\ and\
  \citenamefont {Pietronero}}]{GabrielliBook}%
  \BibitemOpen
  \bibfield  {author} {\bibinfo {author} {\bibfnamefont {A.}~\bibnamefont
  {Gabrielli}}, \bibinfo {author} {\bibfnamefont {F.~S.}\ \bibnamefont
  {Labini}}, \bibinfo {author} {\bibfnamefont {M.}~\bibnamefont {Joyce}},\ and\
  \bibinfo {author} {\bibfnamefont {L.}~\bibnamefont {Pietronero}},\
  }\href@noop {} {\emph {\bibinfo {title} {Statistical Physics for Cosmic
  Structures}}}\ (\bibinfo  {publisher} {Springer Science \& Business Media},\
  \bibinfo {year} {2006})\BibitemShut {NoStop}%
\bibitem [{\citenamefont {Gefen}\ \emph {et~al.}(1983)\citenamefont {Gefen},
  \citenamefont {Meir}, \citenamefont {Mandelbrot},\ and\ \citenamefont
  {Aharony}}]{Gefen1983}%
  \BibitemOpen
  \bibfield  {author} {\bibinfo {author} {\bibfnamefont {Y.}~\bibnamefont
  {Gefen}}, \bibinfo {author} {\bibfnamefont {Y.}~\bibnamefont {Meir}},
  \bibinfo {author} {\bibfnamefont {B.~B.}\ \bibnamefont {Mandelbrot}},\ and\
  \bibinfo {author} {\bibfnamefont {A.}~\bibnamefont {Aharony}},\ }\bibfield
  {title} {\bibinfo {title} {Geometric implementation of hypercubic lattices
  with noninteger dimensionality by use of low lacunarity fractal lattices},\
  }\href {https://doi.org/10.1103/PhysRevLett.50.145} {\bibfield  {journal}
  {\bibinfo  {journal} {Phys. Rev. Lett.}\ }\textbf {\bibinfo {volume} {50}},\
  \bibinfo {pages} {145} (\bibinfo {year} {1983})}\BibitemShut {NoStop}%
\bibitem [{\citenamefont {Binney}\ \emph {et~al.}(1992)\citenamefont {Binney},
  \citenamefont {Dowrick}, \citenamefont {Fisher},\ and\ \citenamefont
  {Newman}}]{Binney}%
  \BibitemOpen
  \bibfield  {author} {\bibinfo {author} {\bibfnamefont {J.~J.}\ \bibnamefont
  {Binney}}, \bibinfo {author} {\bibfnamefont {N.~J.}\ \bibnamefont {Dowrick}},
  \bibinfo {author} {\bibfnamefont {A.~J.}\ \bibnamefont {Fisher}},\ and\
  \bibinfo {author} {\bibfnamefont {M.~E.}\ \bibnamefont {Newman}},\
  }\href@noop {} {\emph {\bibinfo {title} {The theory of critical phenomena: an
  introduction to the renormalization group}}}\ (\bibinfo  {publisher} {Oxford
  University Press},\ \bibinfo {address} {Oxford},\ \bibinfo {year}
  {1992})\BibitemShut {NoStop}%
\bibitem [{\citenamefont {Amit}\ and\ \citenamefont
  {Martin-Mayor}(2005)}]{Amit}%
  \BibitemOpen
  \bibfield  {author} {\bibinfo {author} {\bibfnamefont {D.~J.}\ \bibnamefont
  {Amit}}\ and\ \bibinfo {author} {\bibfnamefont {V.}~\bibnamefont
  {Martin-Mayor}},\ }\href {https://doi.org/10.1142/5715} {\emph {\bibinfo
  {title} {Field Theory, the Renormalization Group, and Critical Phenomena}}},\
  \bibinfo {edition} {3rd}\ ed.\ (\bibinfo  {publisher} {World Scientific},\
  \bibinfo {address} {Singapore},\ \bibinfo {year} {2005})\BibitemShut
  {NoStop}%
\bibitem [{\citenamefont {Kadanoff}(1971)}]{Kadanoff1971}%
  \BibitemOpen
  \bibfield  {author} {\bibinfo {author} {\bibfnamefont {L.~P.}\ \bibnamefont
  {Kadanoff}},\ }in\ \href@noop {} {\emph {\bibinfo {booktitle} {Proceedings of
  the Enrico Fermi Summer School of Physics, Varenna 1970}}},\ \bibinfo
  {editor} {edited by\ \bibinfo {editor} {\bibfnamefont {M.~S.}\ \bibnamefont
  {Green}}}\ (\bibinfo  {publisher} {Academic Press},\ \bibinfo {address}
  {London and New York},\ \bibinfo {year} {1971})\BibitemShut {NoStop}%
\bibitem [{\citenamefont {Gefen}\ \emph {et~al.}(1980)\citenamefont {Gefen},
  \citenamefont {Mandelbrot},\ and\ \citenamefont {Aharony}}]{Gefen1980}%
  \BibitemOpen
  \bibfield  {author} {\bibinfo {author} {\bibfnamefont {Y.}~\bibnamefont
  {Gefen}}, \bibinfo {author} {\bibfnamefont {B.~B.}\ \bibnamefont
  {Mandelbrot}},\ and\ \bibinfo {author} {\bibfnamefont {A.}~\bibnamefont
  {Aharony}},\ }\bibfield  {title} {\bibinfo {title} {Critical phenomena on
  fractal lattices},\ }\href {https://doi.org/10.1103/PhysRevLett.45.855}
  {\bibfield  {journal} {\bibinfo  {journal} {Phys. Rev. Lett.}\ }\textbf
  {\bibinfo {volume} {45}},\ \bibinfo {pages} {855} (\bibinfo {year}
  {1980})}\BibitemShut {NoStop}%
\end{thebibliography}
%
\end{document}